\documentclass[preprint]{aastex63}
\usepackage{amsmath}
\shorttitle{POSITION AND ELLIPTICITY ANGLES}
\shortauthors{McKINNON}

\begin{document}
\title{Behavior of the Position and Ellipticity Angles at Polarization Mode Transitions in 
Pulsar Radio Emission}
\author{M. M. McKinnon}
\affiliation{National Radio Astronomy Observatory, Socorro, NM \ 87801 \ \ USA}
\correspondingauthor{M. M. McKinnon}
\email{mmckinno@nrao.edu}

% --------------------------------------------------------------------------------------------

\begin{abstract}

Polarization observations of radio pulsars show that abrupt transitions in the polarization 
vector's position angle can be accompanied by large excursions in the vector's ellipticity 
angle, suggesting the vector passes near the right or left circular pole of the Poincar\'e 
sphere. The behavior of the angles can be explained by a transition in dominance of the
orthogonal polarization modes or a vector rotation caused by a change in the phase difference 
between the modes. Four polarization models are examined to quantify and understand the 
behavior of the angles at a mode transition: coherent polarization modes, partially coherent 
modes, incoherent modes with nonorthogonal polarization vectors, and incoherent orthogonal 
modes with an elliptically polarized emission component. In all four models, the trajectory 
of the mode transition on the Poincar\'e sphere follows the geodesic that connects the 
orientations of the mode polarization vectors. The results from the models can be similar, 
indicating that the interpretation of an observed transition within the context of a 
particular model is not necessarily unique. The polarization fraction of the emission and 
the average ellipticity angle depend upon the statistical character of the mode intensity 
fluctuations. The polarization fraction increases as the fluctuations increase. The excursion 
in ellipticity angle can be large when the mode intensities are quasi-stable and is suppressed 
when the intensity fluctuations are large.

\end{abstract}

% --------------------------------------------------------------------------------------------

\section{INTRODUCTION}

Polarization observations of radio pulsars show that the position angle (PA) of the emission's
polarization vector can change discontinuously by $90^\circ$, indicating the emission is 
comprised of two modes of orthogonal polarization (OPMs; Manchester, Taylor, \& Huguenin 1975;
Cordes, Rankin, \& Backer 1978). The modes are generally attributed to the natural modes 
of wave propagation in the pulsar magnetosphere (Melrose 1979; Allen \& Melrose 1982; Barnard
\& Arons 1986). Statistical summaries of single-pulse polarization observations (e.g., Backer 
\& Rankin 1980; Stinebring et al. 1984, hereafter S84) show the PAs of the two modes can follow 
parallel tracks across the pulse that are offset from one another by about $90^\circ$, with both 
tracks following the classic rotating vector model (RVM) of Radhakrishnan \& Cooke (1969). 
However, the observations also show that angular separations between the two mode peaks in PA 
histograms can be less than $90^\circ$. Backer \& Rankin (1980) noted that approximately half 
of the stars they observed at 430 MHz had mode PA separations of less than $90^\circ$. S84 
observed many of the same stars at 1404 MHz and found the PA separations were closer to 
$90^\circ$, suggesting the PA separations are frequency-dependent. Additionally, the observed 
shape and extent of a PA transition can deviate from what is expected from a change in the 
dominant polarization mode. The transition is instantaneous and discontinuous over $90^\circ$ 
if the OPMs are incoherent, but many transitions change gradually with pulse longitude, and 
the total change in PA over the duration of the transition can be less than $90^\circ$ (e.g., 
see PSR B0809+74 in Figure 2 of Edwards (2004)). To explain the observed behavior of the PA, 
S84 suggested the emission is comprised of incoherent OPMs and an independent, linearly 
polarized component.

More recent polarization observations of single pulses have reported measurements of the 
emission's ellipticity angle (EA). Some changes in the EA observed at PA transitions are 
unusual and also deviate from what is expected from a change in the dominant polarization 
mode. Ideally, the EA should change discontinuously by $\Delta\chi=2\chi_o$ at an incoherent 
OPM transition, where $\chi_o$ is the EA of one polarization mode and $-\chi_o$ is the EA 
of the other mode. The observed change in the EA, however, can be large, occasionally 
approaching $\Delta\chi=\pm 45^\circ$. For example, the observation of PSR B0329+54 by 
Edwards \& Stappers (2004, hereafter ES04) reveals a large excursion in the EA at the 
pulsar's leading outrider (see their Figure 1). The mean EA abruptly decreases from 
$\chi\simeq 0^\circ$ to $\chi\simeq -35^\circ$ via an asymmetric discontinuity, indicating 
the polarization vector passes near the left circular pole of the Poincar\'e sphere. The 
mean PA at the EA excursion changes discontinuously by about $90^\circ$, as one might 
expect at a transition between incoherent OPMs. Other PA transitions on the leading 
edge of the pulsar's central component and in its trailing outrider are more gradual 
in pulse longitude, and the total change in PA over the transitions appears to fall 
short of the expected $90^\circ$. These PA transitions are generally accompanied by 
inflections in the EA. The EA inflection points are not as prominent or distinct as the 
feature in the pulsar's leading outrider, and the total change in EA at the transitions is 
$\Delta\chi<45^\circ$. In the pulsar's central component, the average EA forms a 
broad asymmetric feature that extends toward the right circular pole of the Poincar\'e 
sphere. The average PA changes gradually at this location, but does not bridge 
the well-defined PA traces of the individual polarization modes. In Figure 2 of ES04, 
individual samples of the PA and EA form a partial annulus in one hemisphere of the 
Poincar\'e sphere and a compact ellipse at the center of the other hemisphere. To 
explain the abrupt EA excursion observed in the pulsar, ES04 suggested the emission is 
comprised of incoherent OPMs and an independent, circularly polarized component. They 
also suggested the EA excursion could be caused by incoherent modes with polarization 
vectors that are not strictly orthogonal. Departures from orthogonality could arise 
from differential refraction of the polarization modes (Cheng \& Ruderman 1979; Barnard 
\& Arons 1986; Petrova 2001; McKinnon 2003, hereafter M03; Oswald et al. 2023b, 
hereafter OKJ). 

Dyks (2020, hereafter D20) noted the polarization modes may be coherent and 
recognized that an EA excursion could be produced by a rotation of the polarization 
vector about the Poincar\'e sphere. The rotation is caused by a change in the phase 
offset between the polarization modes. For some vector geometries and rotations, the 
associated change in PA is gradual and its total change is less than $90^\circ$. Dyks, 
Weltevrede, \& Ilie (2021, hereafter DWI) documented the coherent mode model and showed 
that an EA excursion could also be produced by altering the ratio of the amplitudes of 
the modes' electric fields. The trajectory of this mode transition follows a great circle 
on the Poincar\'e sphere that is offset from the left or right circular pole by an angle 
determined by the phase offset between the modes. The PA in this case changes gradually 
over $90^\circ$. 

Oswald et al. (2023a) conducted a comprehensive analysis of 271 pulsars to identify 
associations between their polarization properties. They found links between PA behavior 
that deviated from the RVM, the frequency evolution of the polarization, and the presence 
of circular polarization features. They suggested the links could be explained by 
propagation effects in the magnetosphere and proposed that the polarization modes were 
partially coherent. OKJ documented the partially coherent model and used it to show that 
observed changes in the EA were consistent with a vector rotation in two pulsars and a 
change in the ratio of mode intensities in another.

In summary, a variety of models have been proposed to explain the behavior of the 
polarization angles at a mode transition. An in-depth examination of the models is 
warranted, to quantify the predicted behavior of the angles and to ascertain the 
applicability of the models. Four polarization models are examined and compared in 
this paper: coherent polarization modes (DWI), partially coherent modes (OKJ), 
incoherent modes with nonorthogonal polarization vectors (M03, ES04), and a combination 
of incoherent OPMs with an elliptically polarized emission component (S84, ES04). 
Henceforth, the models are designated, respectively, as the COH model, the PCOH model, 
the NPM model, and the EPC model. The NPM and EPC models are variants of a statistical 
model of pulsar polarization developed by McKinnon \& Stinebring (1998, hereafter MS).

The paper is organized as follows. The basis for evaluating the models is outlined in 
Section~\ref{sec:basis}. The NPM model is presented in Section~\ref{sec:NPM}, and the 
EPC model is presented in Section~\ref{sec:EPC}. The COH and PCOH models are 
reviewed in Section~\ref{sec:coherent}. The behaviors of the PA, EA, and polarization 
fraction at a mode transition are derived in each case. The properties of the models are 
compared in Section~\ref{sec:compare}. The polarization fraction and average EA depend 
upon the statistical character of the mode intensity fluctuations. The effects of 
different types of mode intensity fluctuations upon the polarization fraction and 
average EA are illustrated in Section~\ref{sec:fluctuate}. The NPM model is used in 
both a numerical simulation of a mode transition and the derivation of the eigenvalues 
and eigenvectors of the Stokes QUV covariance matrix. Section~\ref{sec:discuss} reviews 
the change in EA caused by a rotation of the polarization vector and suggests potential 
applications of the models to select pulsars. Summary comments are listed in 
Section~\ref{sec:summary}. The appendices provide supporting information for the analysis. 
Appendix~\ref{sec:EAGE} lists the equations for the measured EA when the intensities of 
the incoherent OPMs are Erlang or Gaussian random variables (RVs). The eigenvectors 
determined from the NPM model are compared to the mode polarization vectors in 
Appendix ~\ref{sec:eigenvec}.

% --------------------------------------------------------------------------------------------

\section{Definitions and Basis of Model Evaluation}
\label{sec:basis}

For the purpose of evaluating the models, a mode transition is defined as a change 
in the relative mean intensities of the polarization modes that results in a change in the 
dominant mode. The mode transition causes the PA to change by $\Delta\psi\le\pi/2$. A 
rotation of the polarization vector can also change the PA and EA, but it is not a mode 
transition as defined here (see Section~\ref{sec:rotate}). A mode transition so defined 
can be described in more detail with a simple model of incoherent OPMs. The MS statistical 
model of pulsar polarization assumes the emission consists of two independent, 
simultaneously occurring, completely polarized orthogonal modes. The model treats the mode 
intensities as RVs, $X_A$ and $X_B$, to account for the emission's variability and the 
polarization's tendency to switch randomly between orthogonally polarized states. With 
these assumptions, the intensity of the combined radiation is the sum of the RVs, and its 
polarization is related to their difference. The instantaneous values of the Stokes 
parameters I, Q, U, and V at a given pulse longitude are (e.g., M24):

\begin{equation}
I = X_A + X_B
\label{eqn:I}
\end{equation}

\begin{equation}
Q = \cos(2\psi_o)\cos(2\chi_o)(X_A - X_B)
\end{equation}

\begin{equation}
U = \sin(2\psi_o)\cos(2\chi_o)(X_A - X_B)
\end{equation}

\begin{equation}
V = \sin(2\chi_o)(X_A - X_B),
\label{eqn:V}
\end{equation}

\noindent where $\psi_o$ is the PA of the mode A polarization vector and $\chi_o$ is its 
EA. A mode transition in a pulsar's average profile is presumably caused by changes in 
the physical processes that excite the two modes. These changes imply the ratio of the 
mode mean intensities, $M=\langle X_A\rangle/\langle X_B\rangle$, changes with pulse longitude 
to effect the transition. McKinnon (2022, 2024; hereafter M22, M24) introduced a parameter, 
$m$, that is related to the polarization fraction of the radiation. It is defined by

\begin{equation}
m = \frac{\langle X_A\rangle - \langle X_B\rangle}{\langle X_A\rangle + \langle X_B\rangle}
  = \frac{M-1}{M+1}, \qquad -1\le m\le 1.
\label{eqn:m}
\end{equation}

\noindent The radiation is comprised solely of mode A when $m=1$ ($M=\infty$) and is comprised 
solely of mode B when $m=-1$ ($M=0$). The mode transition occurs at $m=0$ ($M=1$). Changes in 
$m$, or $M$, with pulse longitude drive the transition between the modes. Since $m$ must change 
sufficiently to complete the transition, it may be viewed as a proxy for pulse longitude over 
the duration of the transition (M24). Although the parameter $m$ originated in an incoherent 
OPM model, it is applicable in the COH and PCOH models, as shown in Section~\ref{sec:coherent}.

The measured PA is related to the azimuth of the polarization vector in the Poincar\'e sphere 
and is defined by

\begin{equation}
\psi = \frac{1}{2}\arctan{\left(\frac{\langle U\rangle}{\langle Q\rangle}\right)},
\label{eqn:PAdefn}
\end{equation}

\noindent where the angular brackets denote an average over multiple pulses at a given pulse 
longitude. A transition in the PA of $\Delta\psi=\pi/2$ occurs when the mode mean intensities 
are equal to one another, $\langle X_A\rangle = \langle X_B\rangle$ (i.e., when 
$\langle Q\rangle=0$). The measured EA is related to the latitude of the polarization 
vector and is defined by

\begin{equation}
\chi = \frac{1}{2}\arctan{\left(\frac{\langle V\rangle}{\langle L\rangle}\right)},
\label{eqn:chi}
\end{equation}

\noindent where $\langle L\rangle = \langle (Q^2 + U^2)^{1/2}\rangle$ is the mean linear 
polarization. Both the PA and EA are determined by ratios of the Stokes parameters, and since 
the Stokes parameters are determined by the mode intensities, the PA and EA are functions of $m$. 
The polarization models can then be compared against one another by comparing the variations 
of their PAs and EAs with $m$. The models are characterized by multiple parameters, but $m$ 
is the only parameter that is common to all four models. The analysis assumes the other 
parameters remain fixed over the duration of the mode transition.

A direct comparison between the models can be complicated by different geometries assumed 
for the orientations of the mode polarization vectors. Fortunately, the vector geometries 
in MS, DWI, and OKJ are the same. The mode vectors reside along the Stokes Q-axis of the 
Poincar\'e sphere in each model. Therefore, the equations for the Stokes parameters in the 
models can be compared directly without further modification, apart from a normalization for 
consistency in units between models.

% -----------------------------------------------------------------------------------------

\section{Incoherent Modes with Nonorthogonal Polarization Vectors}
\label{sec:NPM}

The NPM model is the MS statistical model modified to account for mode polarization vectors 
that are not strictly orthogonal. Departures from orthogonality in linear and circular
polarization are evaluated separately prior to deriving a general result.

% ---------------------------------------------------------------------------------------------

\begin{figure}
\plotone{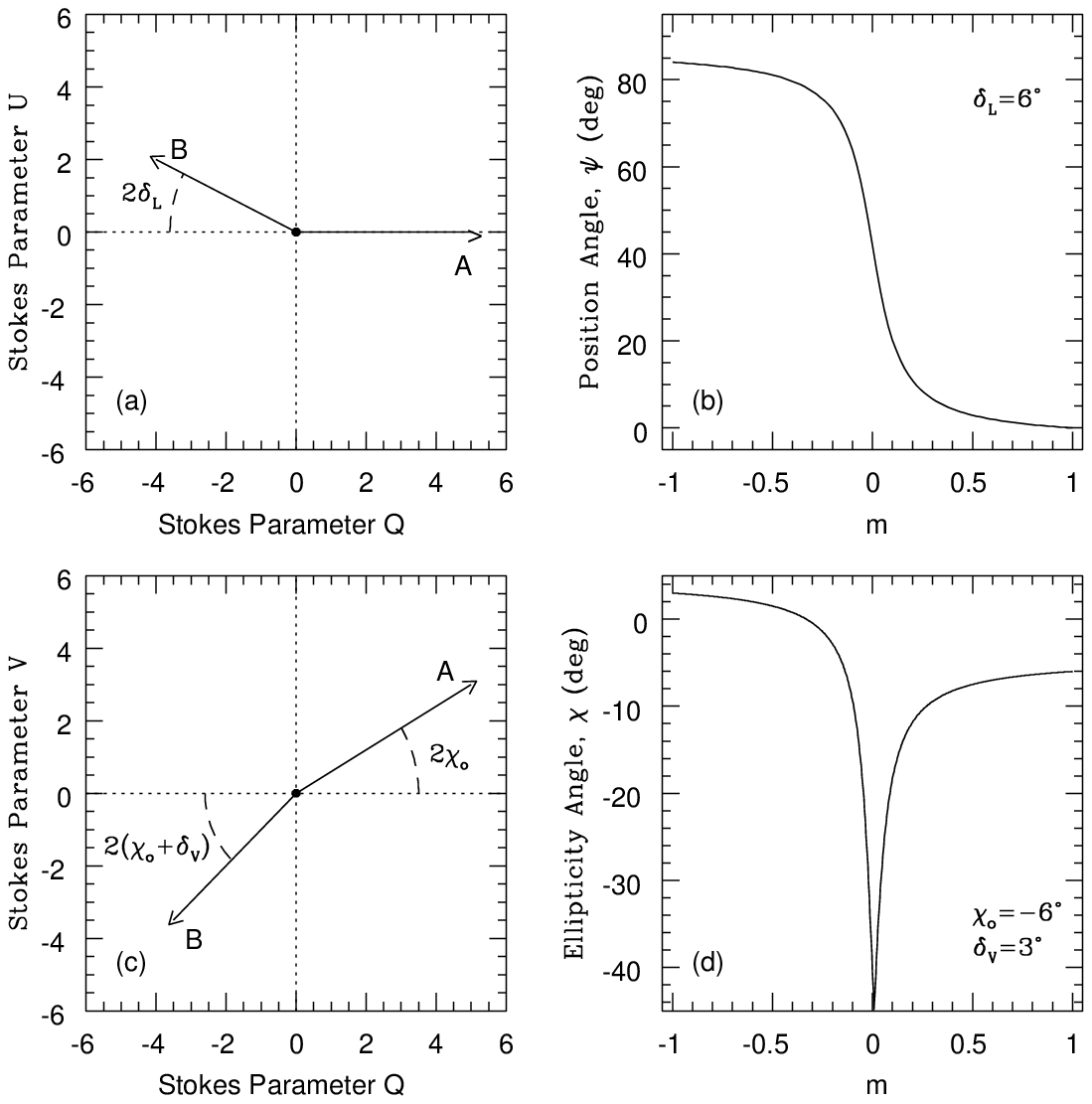}
\caption{Geometry of the mode polarization vectors when they are not orthogonal, and the 
resulting change in the PA and EA at a transition between the polarization modes. Panel (a) 
shows the geometry of the vectors when their linear components are not orthogonal. Panel (b) 
shows the resulting change in the PA at a mode transition (Equation~\ref{eqn:PA}) when 
$\delta_L=6^\circ$. Panel (c) shows the geometry of the vectors when their circular components 
are not orthogonal. Panel (d) shows the resulting change in the EA at a mode transition 
(Equation~\ref{eqn:EA}) when $\chi_o=-6^\circ$ and $\delta_V=3^\circ$.}
\label{fig:nonorthogonal}
\end{figure}

\subsection{Change in the PA at an Incoherent NPM Transition}
\label{sec:PA}

The geometry used to determine how the measured PA is affected by a nonorthogonality in the 
linear component of the mode polarization vectors is shown in panel (a) of 
Figure~\ref{fig:nonorthogonal}. The geometry assumes the EAs of both modes are equal to zero,
so that the vectors reside in the Q-U plane of the Poincar\'e sphere. The PA of the mode A 
vector is $\psi_A=0$, and the PA of the mode B vector is $\psi_B=\pi/2-\delta_L$, where the 
constant angle $\delta_L$ represents the departure from orthogonality in linear polarization. 
Following the analysis originally outlined in M03, the mean values of the Stokes parameters 
Q and U from the geometry in the figure are

\begin{equation}
\langle U\rangle = \langle X_B\rangle\sin(2\delta_L),
\end{equation}

\begin{equation}
\langle Q\rangle = \langle X_A\rangle - \langle X_B\rangle\cos(2\delta_L).
\end{equation}

\noindent By substituting the ratio of the mode mean intensities, 
$M=\langle X_A\rangle/\langle X_B\rangle$, with the parameter $m$ as defined by 
Equation~\ref{eqn:m}, the PA can be shown to vary with $m$ according to 

\begin{equation}
\psi(m) = \frac{1}{2}\arctan\left[\frac{\tan\delta_L(1-m)}{\tan^2\delta_L + m}\right].
\label{eqn:PA}
\end{equation}

\noindent Equation~\ref{eqn:PA} is shown in panel (b) of Figure~\ref{fig:nonorthogonal}. 
The transition occurs at $m_t=-\tan^2\delta_L$, where the PA is $\psi=\pi/4$. The PA is 
antisymmetric about $m=0$, where $\psi=\pi/4-\delta_L/2$. The total change in PA 
across the NPM transition is $\Delta\psi=\pi/2-\delta_L$. 

The rate at which the PA changes with $m$ is 

\begin{equation}
\frac{d\psi}{dm} = -\frac{1}{2}\frac{\tan\delta_L}{(\tan^2\delta_L + m^2)}.
\label{eqn:dPAdm}
\end{equation}

\noindent The rate is symmetric about $m=0$, where the PA changes at the maximum rate of 
$-1/(2\tan\delta_L)$. The rate at the mode transition is slightly less, at 
$-\cos^2\delta_L/(2\tan\delta_L)$. The rates at $m=\pm 1$ are both equal to $-\sin(2\delta_L)/4$.

% ---------------------------------------------------------------------------------------------

\subsection{Change in the EA at an Incoherent NPM Transition}
\label{sec:EA}

The geometry used to determine how the measured EA is affected by a nonorthogonality in the 
circular component of the polarization vectors is shown in panel (c) of 
Figure~\ref{fig:nonorthogonal}. The geometry assumes the PAs of modes A and B are $0$ and 
$\pi/2$, respectively, so that the mode vectors reside in the Q-V plane of the Poincar\'e 
sphere. The EA of the mode A vector is $\chi_o$, and the EA of the mode B vector is 
-($\chi_o+\delta_V$), where the constant angle $\delta_V$ represents the departure from 
orthogonality in circular polarization. Assuming the mode intensities are fixed at the values 
A and B for a given pulse longitude, the circular and linear polarization resulting from the 
geometry shown in the figure are 

\begin{equation}
V = A\sin(2\chi_o)-B\sin[2(\chi_o+\delta_V)],
\label{eqn:VEA}
\end{equation}

\begin{equation}
L = |Q| = |A\cos(2\chi_o)-B\cos[2(\chi_o+\delta_V)]|.
\end{equation}

\noindent By further assuming the mode intensities can vary between pulse longitudes, and 
substituting the ratio of the fixed mode intensities, $M=A/B$, with the parameter $m$, the 
EA can be shown to vary with $m$ across the NPM transition according to

\begin{equation}
\chi(m) = \frac{1}{2}\arctan{\left[\frac{m\tan(2\chi_o+\delta_V)-\tan\delta_V}
          {|\tan(\delta_V)\tan(2\chi_o+\delta_V) + m|}\right]}.
\label{eqn:EA}
\end{equation}

\noindent Equation~\ref{eqn:EA} is shown in panel (d) of Figure~\ref{fig:nonorthogonal}. 
The EA transition occurs at $m_t=-\tan(\delta_V)\tan(2\chi_o+\delta_V)$, where the EA is 
$\chi=-\pi/4$. The polarization vector passes through the left circular pole of the Poincar\'e 
sphere at that value of $m$, because the nonorthogonality has introduced a negative component 
to $V$ that dominates the polarization. The EA excursion is asymmetric about the transition, 
being broader to the right of the transition than to the left of it. The width, $w$, of the 
discontinuity at $\chi=-\pi/8$, where $\tan(2\chi)=-1$, is

\begin{equation}
w=\frac{2\tan\delta_V}{\cos[2(2\chi_o+\delta_V)]}.
\label{eqn:width}
\end{equation}

\noindent The excursion widens as either or both of $\delta_V$ or $\chi_o$ increase. The 
circular polarization fraction, $\bar{\rm V}$, attributable to the nonorthogonality can 
be estimated by setting the mode-intrinsic EA to zero ($\chi_o=0$):

\begin{equation}
\bar{\rm V} = \frac{V}{I}
            = -\frac{B\sin(2\delta_V)}{A + B} = -\sin(2\delta_V)\frac{(1-m)}{2} 
\end{equation}

\noindent The circular polarization fraction varies linearly with $m$ across the 
transition, and is equal to $-\tan\delta_V$ at the transition. 

% -----------------------------------------------------------------------------------------------

\subsection{General Result}
\label{sec:general}

The vector geometries shown in Figure~\ref{fig:nonorthogonal} can be used to derive the changes
in the PA and EA at a mode transition when the departures from mode orthogonality occur in both 
linear and circular polarization. Assuming $\psi_A=0$, $\psi_B=\pi/2-\delta_L$, $\chi_A=0$, and 
$\chi_B=+\delta_V$, the dependence of the average PA upon $m$ is

\begin{equation}
\psi(m) = \frac{1}{2}\arctan\left[\frac{K_3(1 - m)}{K_1 + mK_2}\right],
\label{eqn:PAgen}
\end{equation}

\noindent where $K_1$, $K_2$, and $K_3$ are constants given by 
$K_1=1-\cos(2\delta_V)\cos(2\delta_L)$, $K_2=1+\cos(2\delta_V)\cos(2\delta_L)$, and 
$K_3=\cos(2\delta_V)\sin(2\delta_L)$. The value of $m$ at the transition is 

\begin{equation}
m_t = -\frac{K_1}{K_2} = -\frac{1-\cos(2\delta_V)\cos(2\delta_L)}{1+\cos(2\delta_V)\cos(2\delta_L)}.
\label{eqn:mx}
\end{equation}

\noindent The average EA varies with $m$ according to

\begin{equation}
\chi(m) = \frac{1}{2}\arctan{\left\{\frac{\tan(2\delta_V)(1-m)}
          {[2(K_1 + m^2K_2)(1 + \tan^2(2\delta_V)) - (1-m)^2\tan^2(2\delta_V)]^{1/2}}\right\}}.
\label{eqn:EAgen}
\end{equation}

\begin{figure}
\plotone{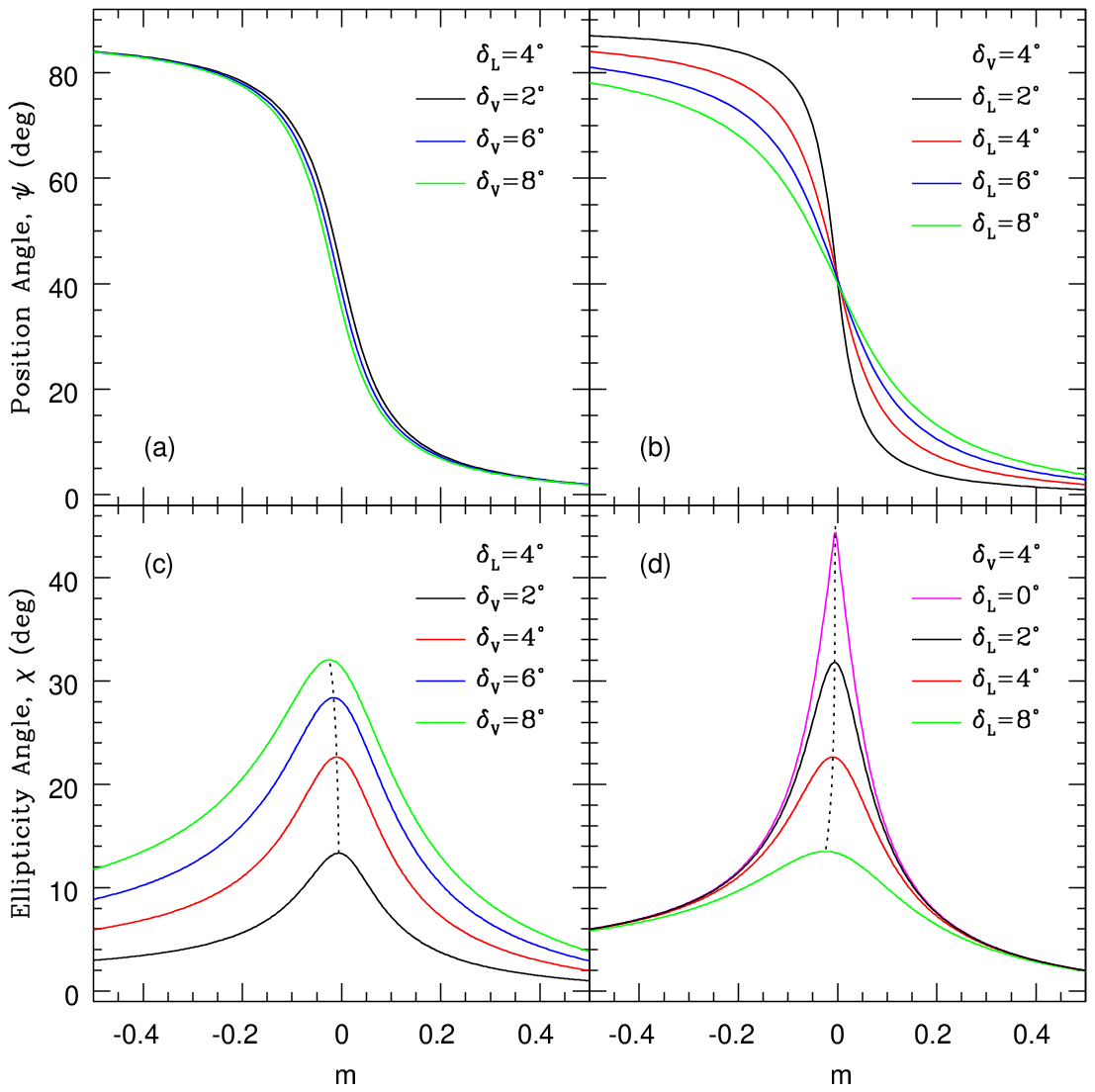}
\caption{Changes in the PA and EA over a transition between NPMs. Panels (a) and (c) in the 
left column of the figure show changes in the PA and EA, respectively, for different values 
of $\delta_V$ when $\delta_L$ is held constant. Panels (b) and (d) in the right column show 
changes in the PA and EA, respectively, for different values of $\delta_L$ when $\delta_V$ 
is held constant. The dotted black lines in panels (c) and (d) connect the vertices of the 
transitions' geodesics (Equation~\ref{eqn:geodesic}).}
\label{fig:EAPA}
\end{figure}

\noindent Examples of Equations~\ref{eqn:PAgen} and~\ref{eqn:EAgen} are shown in 
Figure~\ref{fig:EAPA}. The EA excursion is asymmetric about the transition point, $m_t$.
The slope of the PA transition decreases with increasing values of $\delta_L$. The excursion in 
the EA increases with increasing $\delta_V$ and decreasing $\delta_L$. The EA reaches its 
peak value of $\chi=\pi/4$ when $\delta_L=0$. The dotted lines in the lower row of panels 
denote the peak EA at each mode transition and are discussed in Section~\ref{sec:trajectory}. 

If the mode intensities are fixed at a given pulse longitude, but can vary with longitude, 
the polarization fraction of the NPM model as a function of $m$ is

\begin{equation}
p(m) = (Q^2+U^2+V^2)^{1/2}/I = [(K_1 + m^2K_2)/2]^{1/2}.
\label{eqn:pn}
\end{equation}

\noindent The fraction is symmetric about $m=0$, where its minimum value is 
$p_m = \sqrt{K_1/2}$. The minimum is offset from the mode transition at $m_t$, where 
the polarization fraction is $p(m_t)=\sqrt{K_1/K_2}$. For incoherent OPMs ($\delta_L=\delta_V=0$), 
the polarization fraction is $p(m)=|m|$.

The derivation of Equation~\ref{eqn:PAgen} for the average PA assumes the PA of mode A remains
constant at zero across the mode transition. If the PA of mode A is constant at some other
value ($\psi_o$), or if it varies linearly, and slowly, with $m$ across the mode 
transition (e.g., $\psi_o(m)=K_0m$), the PA is

\begin{equation}
\psi(m) = \frac{1}{2}\arctan\left[\frac{\tan(2\psi_o(m))(K_1 + mK_2) + K_3(1 - m)}
          {(K_1 + mK_2) - \tan(2\psi_o(m))K_3(1 - m)}\right].
\label{eqn:PAK}
\end{equation}

\noindent The PA given by Equation~\ref{eqn:PAK} varies continuously with $m$ as long as 
the total change in PA across the transition is $\Delta\psi<\pi/2$. This constraint restricts 
values of $K_0$ to $\vert K_0\vert < \pi/4$.

% --------------------------------------------------------------------------------------------

\subsection{Trajectory of a Mode Transition on the Poincar\'e Sphere}
\label{sec:trajectory}

\begin{figure}
\plotone{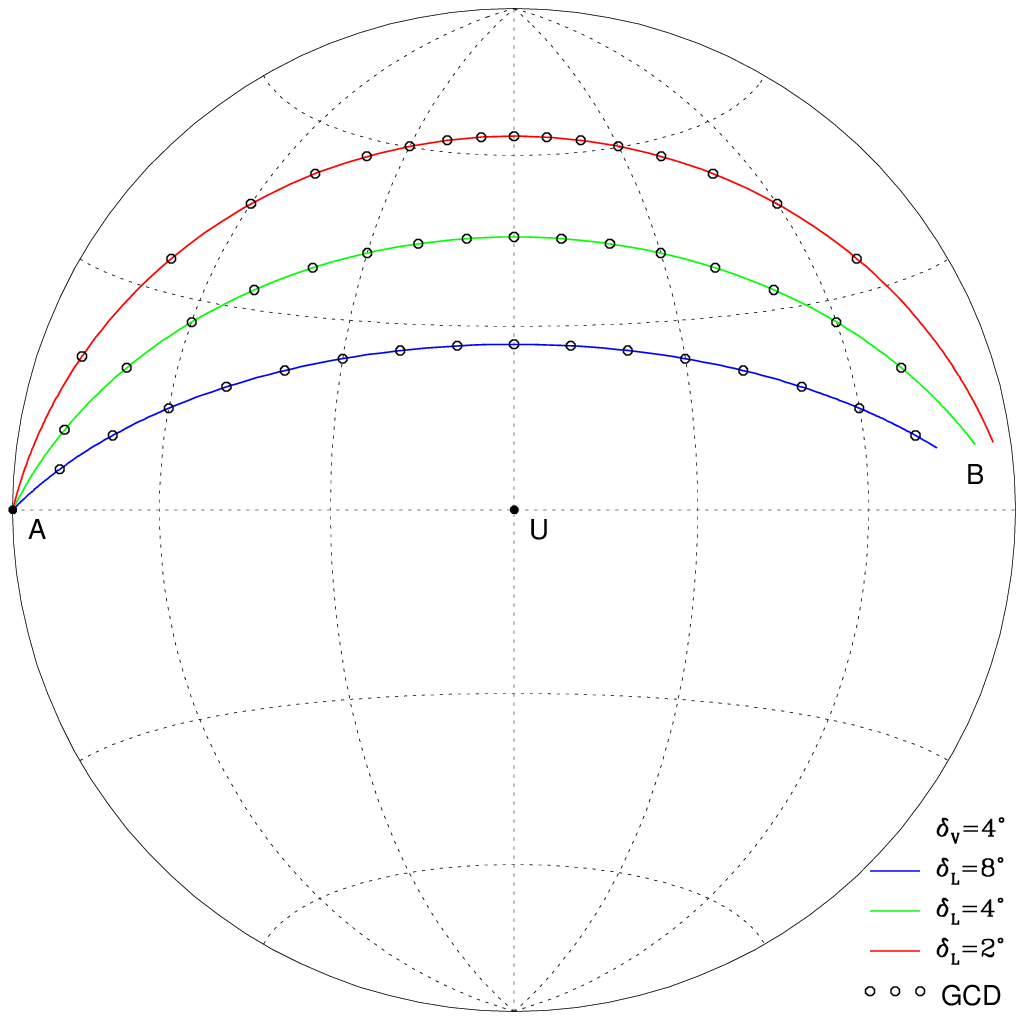}
\caption{Example trajectories of NPM transitions on the Poincar\'e sphere. The trajectories
(colored lines) are the projections of $\psi$ and $\chi$ as determined from 
Equations~\ref{eqn:PAgen} and~\ref{eqn:EAgen}, respectively. The values of $\delta_V$ and 
$\delta_L$ used to calculate the trajectories are annotated in the figure. The open circles
trace the geodesics or GCDs between the orientations of the mode polarization vectors, as 
calculated from Equation~\ref{eqn:geodesic}. The trajectories follow the geodesics. The 
view of the sphere is along its U-axis, such that mode A is on the sphere equator at the 
left, and mode B is at the terminus of each trajectory on the right.}
\label{fig:arcs}
\end{figure}

Projections of the PA and EA, as calculated from Equations~\ref{eqn:PAgen} and~\ref{eqn:EAgen},
on the Poincar\'e sphere are shown by the colored lines in Figure~\ref{fig:arcs}. The view of the 
sphere is along its U-axis ($\psi=\pi/4$), with mode A on the sphere equator to the left and mode 
B on the right. The trajectories of the mode transitions were calculated for different values of 
$\delta_L$, with $\delta_V$ held constant at $\delta_V=4^\circ$. For each example in the figure, 
the peak value of the EA occurs at $\psi=\pi/4$ and increases as the value of $\delta_L$ decreases. 
The trajectories follow the geodesics between the orientations of the mode polarization vectors. 
The orientations of the mode vectors on the sphere and the sphere's origin define a unique plane 
that intersects the origin. The intersection between the plane and the surface of the sphere is
a great circle. The shortest distance between the orientations of the mode vectors on the sphere
lies along this great circle and is called the geodesic or great circle distance (GCD; e.g., 
Kells, Kern, \& Bland 1940). The relationship between the PA and EA over the trajectory is given 
by the equation for the great circle defined by the mode vectors. For the geometry used to 
derive Equations~\ref{eqn:PAgen} and~\ref{eqn:EAgen}, the relationship between $\chi$ and $\psi$ 
is 

\begin{equation}
\chi(\psi) = \frac{1}{2}\arctan{\left[\frac{\sin(2\psi)\tan(2\delta_V)}{\sin(2\delta_L)}\right]}.
\label{eqn:geodesic}
\end{equation}

\noindent Equation~\ref{eqn:geodesic} is shown by the open circles in Figure~\ref{fig:arcs} 
for the values of $\delta_L$ and $\delta_V$ annotated in the figure. The figure shows that 
Equation~\ref{eqn:geodesic} replicates the dependence of $\chi$ upon $\psi$, as determined 
independently from Equations~\ref{eqn:PAgen} and~\ref{eqn:EAgen}. The vertex of the geodesic 
(the peak value of $\chi$) occurs at $\psi=\pi/4$ and is large when $\delta_V$ is large and 
$\delta_L$ is small. The vertex is also shown by the black dotted lines in panels (c) and (d) 
of Figure~\ref{fig:EAPA}. The dotted lines connect the peak EAs of each example shown in the 
two panels. When $\delta_V=0$, the transition trajectory traverses the equator of the 
Poincar\'e sphere. When $\delta_L=0$, the trajectory traverses a meridian of the sphere over 
one of its poles. 

The angular extent, $\zeta$, of the geodesic is the angle subtended by the mode polarization 
vectors, 

\begin{equation}
\zeta = \pi - \arccos[\cos(2\delta_V)\cos(2\delta_L)], 
\label{eqn:zeta}
\end{equation}

\noindent which is always less than $\pi$. Since the radius of the Poincar\'e sphere
is equal to $1$, $\zeta$ is also the GCD of the transition trajectory.

% --------------------------------------------------------------------------------------------

\section{Incoherent Orthogonal Modes with an Elliptically Polarized Component}
\label{sec:EPC}

S84 suggested that a total change in PA of less than $\pi/2$ over a mode transition could be 
explained by a combination of incoherent OPMs and an independent, linearly polarized emission 
component. Similarly, ES04 proposed that EA excursions coinciding with PA transitions could 
be explained by a combination of incoherent OPMs and an independent, circularly polarized 
component. Both suggestions can be examined by assuming the independent component is 
elliptically polarized. The physical origin of the EPC is unknown. The derivations of the 
PA and EA as functions of $m$ in this case are very similar to those presented in 
Section~\ref{sec:NPM}. The OPMs are assumed to be linearly polarized, and their vectors are 
assumed to be aligned with the Stokes Q-axis of the Poincar\'e sphere. The intensity of the 
EPC is designated as $\varepsilon$, and its PA and EA are denoted by $\psi_e$ and $\chi_e$, 
respectively. The mode intensities are assumed to be fixed at A and B for a given pulse 
longitude, and the combined intensity of the three components is assumed to be constant and 
equal to $1$, such that $A+B=1-\varepsilon$. The Stokes parameters resulting from these 
assumptions for the case of $\psi_e=\pi/4$ are

\begin{equation}
Q = A - B,
\end{equation}

\begin{equation}
U = \varepsilon\cos(2\chi_e),
\end{equation}

\begin{equation}
V = \varepsilon\sin(2\chi_e).
\end{equation}

\noindent The EPC model is the scenario described by S84 when $\chi_e=0$ and is the 
scenario proposed by ES04 when $\chi_e=\pi/4$. The PA and EA derived from the above 
Stokes parameters are

\begin{equation}
\psi(m) = \frac{1}{2}\arctan{\left[\frac{\varepsilon\cos(2\chi_e)}{m(1-\varepsilon)}\right]},
\label{eqn:PAE}
\end{equation}

\begin{equation}
\chi(m) = \frac{1}{2}\arctan{\Biggl\{\frac{\varepsilon\tan(2\chi_e)}{\left[m^2(1-\varepsilon)^2
         (1+\tan^2(2\chi_e)) + \varepsilon^2\right]^{1/2}}\Biggr\}}.
\label{eqn:EAE}
\end{equation}

\begin{figure}
\plotone{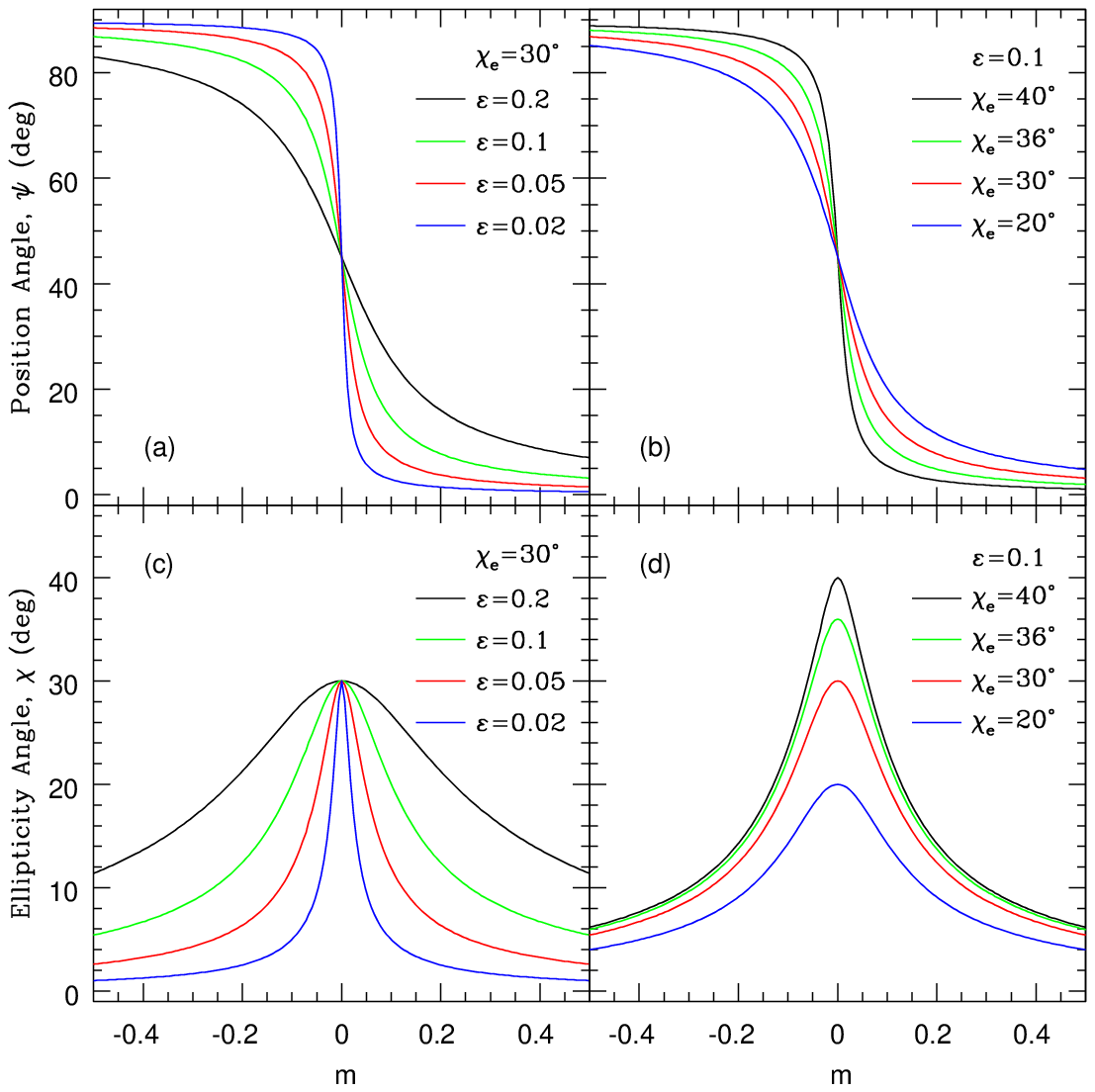}
\caption{Changes in the PA and EA over a mode transition predicted by the EPC model. 
Panels (a) and (c) in the left column show the change in the PA and EA for different values 
of $\varepsilon$ when $\chi_e$ is held constant. Panels (b) and (d) in the right column of 
the figure show the change in PA and EA for different values of $\chi_e$ when $\varepsilon$ 
is held constant.}
\label{fig:EPC}
\end{figure}

\noindent Examples of Equations~\ref{eqn:PAE} and~\ref{eqn:EAE} are shown in 
Figure~\ref{fig:EPC}. The mode transition occurs at $m=0$, where the PA changes at its 
maximum rate of

\begin{equation}
\frac{d\psi}{dm} = -\frac{(1-\varepsilon)}{2\varepsilon\cos(2\chi_e)}.
\end{equation}

\noindent The total change in the PA over the transition is $\Delta\psi=\pi/2-2\psi_s$, 
where $\psi_s$ is given by Equation~\ref{eqn:PAE} with $m=1$. The EA excursion is symmetric 
about $m=0$, where its peak value is $\chi_p=\chi_e$. The full width of the EA excursion at 
half its maximum value is

\begin{equation}
w = \frac{2\varepsilon}{1-\varepsilon}\Biggl(\frac{3-\tan^2\chi_e}{1+\tan^2\chi_e}\Biggr)^{1/2}.
\end{equation}

\noindent The PA rate increases and the EA excursion narrows as $\varepsilon$ decreases 
and as $\chi_e$ increases. 

The trajectory of an EPC mode transition also follows a geodesic on the Poincar\'e 
sphere. The equation for the great circle that defines the trajectory, and thus the 
dependence of the EA upon the PA, is

\begin{equation}
\chi(\psi) = \frac{1}{2}\arctan{\left[\sin(2\psi)\tan(2\chi_e)\right]}.
\end{equation}

\noindent The vertex of the EA excursion occurs at $\psi=\pi/4$. The GCD of the 
trajectory is

\begin{equation}
\zeta = \pi - \arccos{\left(\frac{1-2\varepsilon}{1+2\varepsilon^2-2\varepsilon}\right)}.
\label{eqn:EPCzeta}
\end{equation}

The polarization fraction for the EPC model is

\begin{equation}
p(m) = [m^2(1-\varepsilon)^2 + \varepsilon^2]^{1/2}.
\label{eqn:pe}
\end{equation}

\noindent The polarization fraction is independent of $\chi_e$ and is symmetric about its
minimum value of $p_m=\varepsilon$ at $m=0$.

% --------------------------------------------------------------------------------------------

\section{COHERENT AND PARTIALLY COHERENT MODES}
\label{sec:coherent}

The PCOH model developed by OKJ uses three parameters to define the Stokes parameters 
of the radiation: a coherence factor, $C$; a mode strength ratio, $R$; and a mode phase offset, 
$\eta$ (see their Equation 5). As OKJ note, the emission produced by partially coherent modes 
may be interpreted as a combination of coherent and incoherent components. The polarization 
fraction of the incoherent component is defined by the ratio of OKJ's Stokes parameters Q and 
I. The ratio is independent of $C$ and $\eta$, and is equal, within an immaterial factor of 
-1, to the parameter $m$ given by Equation~\ref{eqn:m}. OKJ's Stokes parameters U, V, and I 
determine the polarization fraction of the coherent component. When the coherence factor is 
$C=0$, the modes are incoherent, and the PCOH model becomes the MS model of incoherent OPMs. 
The mode intensity ratio, $M$, defined in MS is identical to OKJ's mode strength ratio, $R$. 
When $C=1$, the modes are coherent, and the PCOH model is equivalent to the COH model (see 
Equations 13-15 in DWI). The COH model is characterized by two parameters: a mode mixing 
angle, $\psi_{mx}$, and a mode phase lag, $\delta_{ox}$. The phase lag in DWI is the same as 
the mode phase offset in OKJ, $\eta=\delta_{ox}$, and DWI's mixing angle is related to OKJ's 
mode strength ratio by

\begin{equation}
\cos(2\psi_{mx}) = \frac{1-R}{1+R} = -m.
\end{equation}

\begin{figure}
\plotone{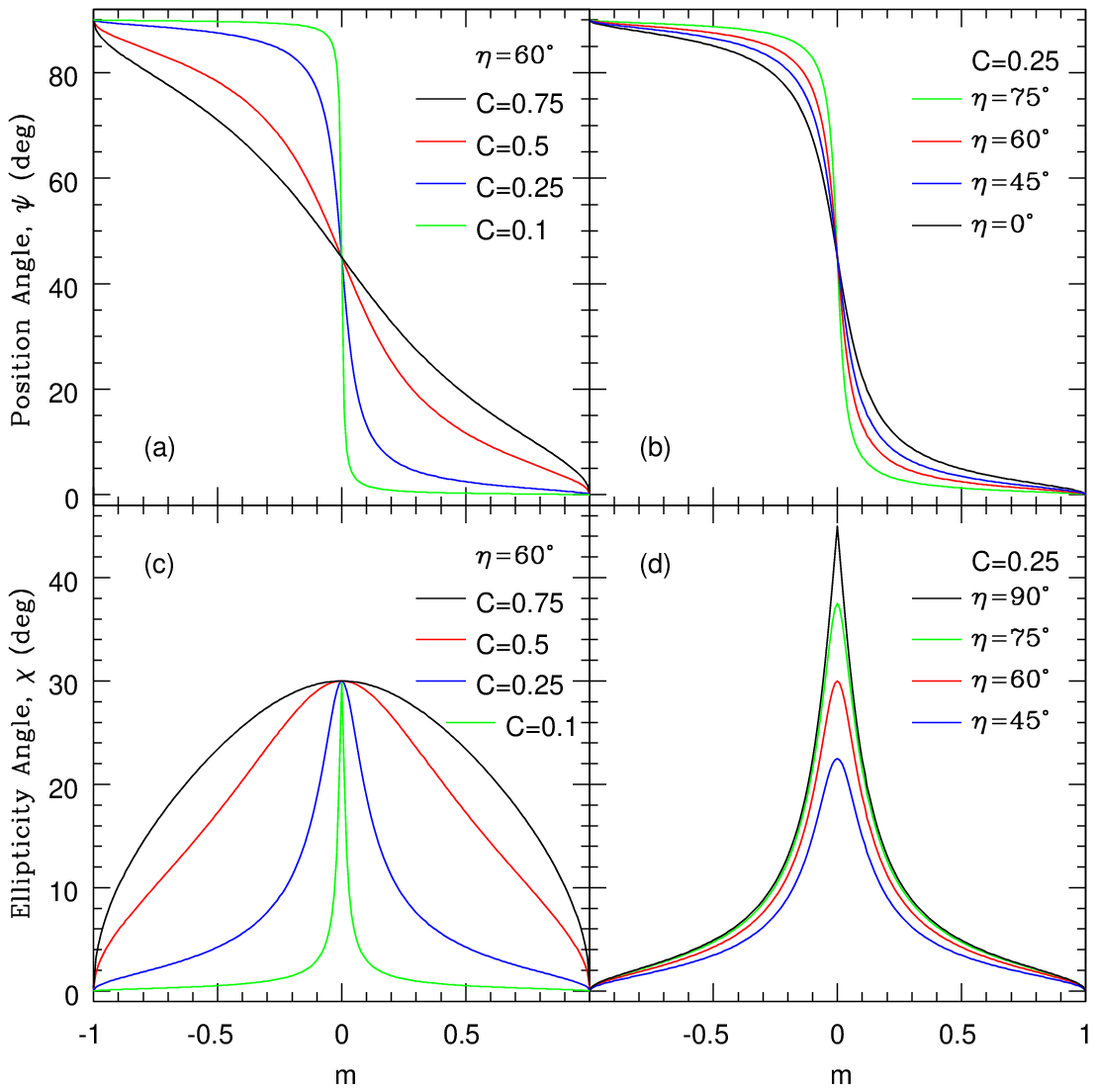}
\caption{Changes in the PA and EA over a transition between partially coherent polarization 
modes. Panels (a) and (c) in the left column of the figure show the change in PA and EA, 
respectively, for different values of $C$ when the phase offset is held constant at 
$\eta=60^\circ$. Panels (b) and (d) in the right column of the figure show the change in PA 
and EA, respectively, for different values of $\eta$ when the coherence factor is held 
constant at $C=0.25$.}
\label{fig:PCOH}
\end{figure}

\noindent The Stokes parameters defined in DWI and OKJ can be used to derive the dependence 
of the PA and EA upon $m$ for fixed values of $C$ and $\eta$. Using OKJ's notation, the PA 
and EA are

\begin{equation}
\psi(m) = \frac{1}{2}\arctan{\left[\frac{\cos\eta(1-m^2)^{1/2}}{t(C)m}\right]},
\label{eqn:PAP}
\end{equation}

\begin{equation}
\chi(m) = \frac{1}{2}\arctan{\Biggl\{\tan\eta\left[\frac{1-m^2}{t^2(C)m^2(1+\tan^2\eta)
        + (1-m^2)}\right]^{1/2}\Biggr\}},
\label{eqn:EAP}
\end{equation}

\noindent where the factor $t(C)=[(1-C)^2+C^2]/C^2$ ranges from $t(1)=1$ to $t(0)=\infty$. 
Examples of Equations~\ref{eqn:PAP} and~\ref{eqn:EAP} are shown in Figure~\ref{fig:PCOH}. 
The mode transition occurs at $m=0$, where the PA changes at its maximum rate of 
$d\psi/dm=-t(C)/(2\cos\eta)$. The EA excursion is symmetric about $m=0$, where its peak 
value is $\chi_p=\eta/2$. The full width of the EA excursion at half its maximum value is 

\begin{equation}
w = 2\Biggl\{\frac{3-\tan^2(\eta/2)}{4+[t^2(C)-1][1+\tan^2(\eta/2)]}\Biggr\}^{1/2}.
\end{equation}

\noindent The PA rate increases and the EA excursion narrows as $\eta$ increases and as 
$C$ decreases.

For coherent polarization modes ($C=t(C)=1$), the PA changes at a maximum rate of 
$d\psi/dm=-1/(2\cos(\eta))$. The PA transition can be gradual or discontinuous, although 
abrupt transitions require $\eta\sim\pi/2$. The EA excursion for the COH model has a 
width of $w=[3-\tan^2(\eta/2)]^{1/2}$, and is generally wider than the excursions produced 
by the PCOH model. The behavior of the polarization angles predicted by the COH model is 
not shown in a separate figure, but may be inferred from Figure~\ref{fig:PCOH} for the 
PCOH model.

The trajectories of the COH and PCOH mode transitions on the Poincar\'e sphere follow great 
circles. The equation for the great circle is

\begin{equation}
\chi(\psi) = \frac{1}{2}\arctan{[\sin(2\psi)\tan\eta]},
\end{equation}

\noindent and it applies to both coherent and partially coherent modes. The GCD of the 
trajectory is $\zeta=\pi$, because the mode vectors reside on antipodal points of the 
Poincar\'e sphere.

The polarization fraction of the PCOH model figures prominently in OKJ's analysis
and interpretation of their observations. The polarization fraction for both the COH and 
PCOH models is

\begin{equation}
p(m) = [m^2(t^2(C)-1) + 1]^{1/2}/t(C).
\label{eqn:pp}
\end{equation}

\noindent The fraction is independent of the mode phase offset, $\eta$, and is symmetric 
about its minimum value of $p_m=1/t(C)$ at $m=0$. For the COH model, the polarization 
fraction remains constant at $p=1$ over the transition's duration.

The application of the PCOH model to a mode transition described here is the same procedure 
used by OKJ to interpret the polarization of PSR J0134-2937 (see their Figure 9). They found 
that values of $C$ and $\eta$ were approximately constant with pulse longitude and showed 
that $R$, which is related to the parameter $m$, varied roughly linearly with longitude. As 
Equation~\ref{eqn:pp} shows, once $C$ is known, the polarization fraction observed over a 
range of longitudes can provide a direct mapping between $m$, or $R$, and longitude.

% ------------------------------------------------------------------------------------------

\section{Model Comparison}
\label{sec:compare}

\begin{deluxetable}{cccccc}
\tablenum{1}
\tablecaption{Comparison of Polarization Model Properties}
\tablehead{\colhead{Model} & \colhead{NPM} & \colhead{EPC} & \colhead{PCOH} & \colhead{COH}
          & \colhead{ICOH}}
\startdata
$\Delta\psi$ & $\pi/2-\delta_L$ & $\pi/2-2\psi_s$ & $\pi/2$ & $\pi/2$ & $\pi/2$ \\
$\tan(2\chi_p)$ & $\tan(2\delta_V)/\sin(2\delta_L)$ & $\tan(2\chi_e)$ & $\tan\eta$ & $\tan\eta$ 
                & 0 \\
GCD, $\zeta$ & $<\pi$ & $<\pi$ & $\pi$ & $\pi$ & $\pi$ \\
$p_m$ & $\sqrt{K_1/2}$ & $\varepsilon$ & $1/t(C)$ & 1 & 0 \\
$N$ & 3 & 3 & 3 & 2 & 1 \\
\enddata
\end{deluxetable}

The properties of the polarization models are compared in Table 1. For completeness, the 
properties of a transition between incoherent OPMs (ICOH) have been included in the table. 
The table entries include the total change in the PA across the mode transition, $\Delta\psi$; 
the tangent of the peak EA, $\chi_p$; the GCD of each mode transition, $\zeta$; the minimum 
polarization fraction, $p_m$; and the number of parameters used to characterize each model, $N$. 
The table entries, combined with Figures~\ref{fig:EAPA},~\ref{fig:EPC}, and~\ref{fig:PCOH}, 
show the models have similarities, as well as differences. The PA transition and EA excursion 
of each model coincide with one another, and their durations are comparable. The PA transitions 
of the COH model tend to be more gradual and its EA excursions are generally broader than their 
counterparts in the other models. The PA transition for each model is antisymmetric about 
$m=0$; however, the transition point for the NPM model, $m_t$, does not coincide with $m=0$. 
The PA changes discontinuously in a transition in the ICOH model, but changes gradually over a 
transition in the other models. The total change in the PA over a transition in the PCOH, COH, 
and ICOH models is $\Delta\psi=\pi/2$ and is less than $\pi/2$ in the NPM and EPC models. In 
all cases except for the ICOH model, the EA can trace an excursion that originates near the 
equator of the Poincar\'e sphere and passes near one of its poles. The ICOH model is not 
capable of producing an EA excursion. The excursion in the EPC, PCOH, and COH models is 
symmetric about a transition, but is asymmetric about a transition in the NPM model. The 
maximum value of the EA within an excursion, $\chi_p$, is determined by the model parameters. 
For each model, the trajectory of a mode transition on the Poincar\'e sphere follows the 
geodesic connecting the orientations of the mode polarization vectors, with the details of the 
trajectory determined by the model's parameters. In the ICOH model, the orientations of the 
mode vectors reside at antipodal points of the Poincar\'e sphere. A unique great circle does 
not connect these orientations, because an infinite number of great circles can connect 
antipodal points on the sphere. The GCDs of the transition trajectories for the PCOH, COH, 
and ICOH models are equal to $\pi$, and are less than $\pi$ in the NPM and EPC models.

Of the models listed in Table 1, only the NPM model is capable of reproducing the asymmetric 
EA excursion observed in the leading outrider of PSR B0329+54. The asymmetric EA excursion 
and offset between $m_t$ and $m=0$ in the model arise because the nonorthogonality of the 
mode polarization vectors is a mechanism that breaks the symmetry of the mode transition. The 
other models do not include a symmetry-breaking mechanism. An asymmetry can be introduced in 
the EPC model by allowing the PA of the EPC, $\psi_e$, to be something other than $\pi/4$, 
which would increase the number of model parameters from $N=3$ to $N=4$.

By construction, the orientation of the EPC's polarization vector in the EPC model is the 
same as that of the coherent component in the PCOH model. However, the two models are not 
the same, because the intensity of the EPC is independent of the mode intensities, whereas 
the intensity of the coherent component in the PCOH model is dependent upon them.

\begin{figure}
\plotone{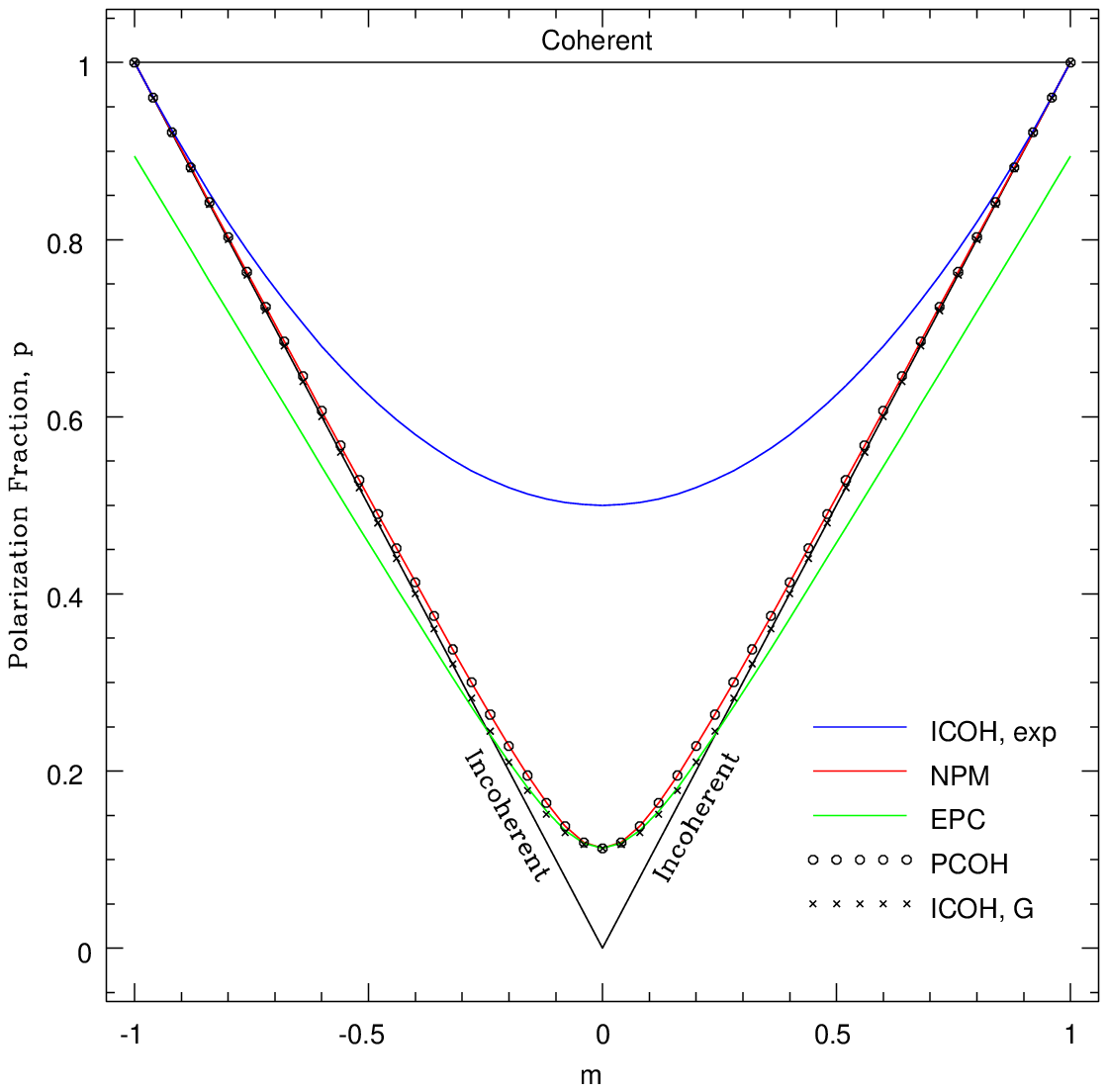}
\caption{Comparison of model polarization fractions, $p(m)$. The red line is $p(m)$ for the 
NPM model (Equation~\ref{eqn:pn} with a cosine product of $\cos(2\delta_V)\cos(2\delta_L)=0.975$), 
and the open circles represent $p(m)$ for the PCOH model (Equation~\ref{eqn:pp} with a coherence 
factor of $C=0.263$). The green line is $p(m)$ for the EPC model (Equation~\ref{eqn:pe} with 
$\varepsilon=0.113$), and the black crosses represent $p(m)$ produced by Gaussian fluctuations 
in incoherent OPM intensities (Equation~\ref{eqn:pg} with a modulation index of $\beta=0.141$). 
The blue line is $p(m)$ produced by exponential fluctuations in incoherent OPM intensities 
(Equation~\ref{eqn:px}). With the exception of the exponential fluctuation case, the minimum 
polarization fraction for each model is $p_m=0.113$.}
\label{fig:polfrac}
\end{figure}

The polarization fractions of the models are compared in Figure~\ref{fig:polfrac}. The 
figure is adopted from Figure 2 of Cheng \& Ruderman (1979) and Figure 5 of M24. For models 
that assume the emission is comprised solely of the polarization modes, the polarization 
fraction is confined to the region defined by the inverted triangle in the figure. The top
horizontal boundary at $p=1$ is formed by coherent polarization modes and the lower 
diagonal boundaries are formed by incoherent OPMs with intensities that do not randomly 
vary. With one exception, the polarization fractions are displayed using the same value of 
the minimum polarization fraction, $p_m=0.113$. The polarization minima increase as $C$ and 
$\varepsilon$ increase and as $\cos(2\delta_V)\cos(2\delta_L)$ decreases. The overall 
polarization fractions retain their hyperbola-like shapes as the model parameters change. 
The figure shows very little difference between the polarization fractions of the NPM, EPC, 
and PCOH models over the range of $m=\pm 0.1$ ($0.82<M<1.22$). In fact, the functional 
forms of the polarization fractions for the NPM and PCOH models (Equations~\ref{eqn:pn} 
and~\ref{eqn:pp}) are mathematically the same, with $t^2(C)=2/K_1$. This, in turn, means the 
physical properties of a mode transition cannot be uniquely determined on the basis of the 
polarization fraction alone. The polarization fraction of the EPC model extends beyond the 
boundary indicated by the inverted triangle, because the model includes an EPC in addition 
to the polarization modes. The polarization fraction at the beginning and end of the EPC 
mode transition is given by Equation~\ref{eqn:pe} with $m=1$. The polarization fraction of 
the COH model remains constant at $p=1$, which is generally inconsistent with what is 
observed in individual pulsars and across the pulsar population (e.g., see Figures 3 and 9 
of OKJ).

% ------------------------------------------------------------------------------------------

\section{Effects of Mode Intensity Fluctuations}
\label{sec:fluctuate}

The preceding analyses have generally assumed the mode intensities do not randomly vary. 
However, the radio emission has long been known to fluctuate (e.g., Ekers \& Moffet 1969; 
Manchester, Taylor, \& Huguenin 1975; Backer \& Rankin 1980; S84), and the observed switching 
between orthogonally polarized states is generally regarded as a stochastic process (Cordes, 
Rankin, \& Backer 1978). A model that hopes to replicate the results of single-pulse polarization 
observations (e.g., S84 and ES04), such as the modulation index of the total intensity, the 
eigenvalues and eigenvectors of the Stokes QUV covariance matrix, and distributions of the PA, 
EA, and fractional polarization, should incorporate intensity fluctuations. MS incorporated the 
fluctuations as a fundamental component in their statistical model of incoherent OPMs.

\subsection{Polarization Fraction from Fluctuations in Incoherent OPM Intensities}

The mode intensity fluctuations contribute to the polarization fraction, and the contribution 
is dependent upon the statistical character of the fluctuations. As shown by Equation 18 of 
M22, the polarization fraction\footnote{M22 and M24 refer to Equations~\ref{eqn:px} 
and~\ref{eqn:pg} as the normalized mean of linear polarization or the fractional linear 
polarization. The equations also represent the polarization fraction, as used here, because
M22 and M24 assumed the modes are completely linearly polarized.} resulting 
from exponential fluctuations in the intensities of incoherent OPMs is

\begin{equation}
p(m) = \frac{M^2+1}{(M+1)^2} = \frac{1 + m^2}{2}.
\label{eqn:px}
\end{equation}

\noindent Equation~\ref{eqn:px} is shown by the blue line in Figure~\ref{fig:polfrac}. Its 
minimum polarization fraction is $p_m=0.5$, demonstrating that the contributions of the 
intensity fluctuations to the polarization fraction can be significant. The modulation index,
$\beta$, varies over the transition from $\beta=0.71$ at $m=0$ to $\beta=1$ at $m=\pm 1$ 
(see Equation 20 and Figure 2 of M24).

For Gaussian fluctuations in mode intensities, the polarization fraction from Equation 45 
of M24 is

\begin{equation}
p(m) = \beta\sqrt{\frac{2}{\pi}}\exp{\left(-\frac{m^2}{2\beta^2}\right)}
     + {\rm erf}\left(\frac{m}{\beta\sqrt{2}}\right)m,
\label{eqn:pg}
\end{equation}

\noindent where $\rm{erf}(x)$ is the error function. The modulation index is constrained 
by $\beta\le 1/(5\sqrt{2})=0.141$ to ensure the mode intensities are non-negative (see Equation 
40 of M24 and its supporting text). For simplicity and illustration purposes, the total 
intensity of the emission, the standard deviation of the mode intensities, and thus the 
modulation index are assumed to be constant across the transition. Equation~\ref{eqn:pg} with 
$\beta=0.141$ is shown by the black crosses in Figure~\ref{fig:polfrac}. Since $\beta=0.141$ is 
the largest it can be for Gaussian intensity fluctuations, the polarization minimum can be no 
larger than $p_m=0.113$. This is the constraint that sets the value of $p_m$ for the model 
comparison in the figure. The polarization fraction produced by Gaussian fluctuations in mode 
intensities replicates that of the EPC model over the range $m=\pm 0.25$. Polarization fractions 
produced by other types of intensity fluctuations can occupy the $m-p$ parameter space delimited 
by the exponential and Gaussian examples shown in the figure (see Figure 5 of M24). The examples 
show that the polarization fraction produced by incoherent OPMs with fluctuating intensities can 
mimic the polarization produced by more complex models that assume the mode intensities do not 
fluctuate. Oswald et al. (2023a) assert that the polarization is always equal to zero at a 
transition between incoherent OPMs and attribute any residual polarization to propagation 
effects. The polarization is equal to zero only when the OPM intensities are equal and do not 
randomly vary.

% ------------------------------------------------------------------------------------------

\begin{figure}
\plotone{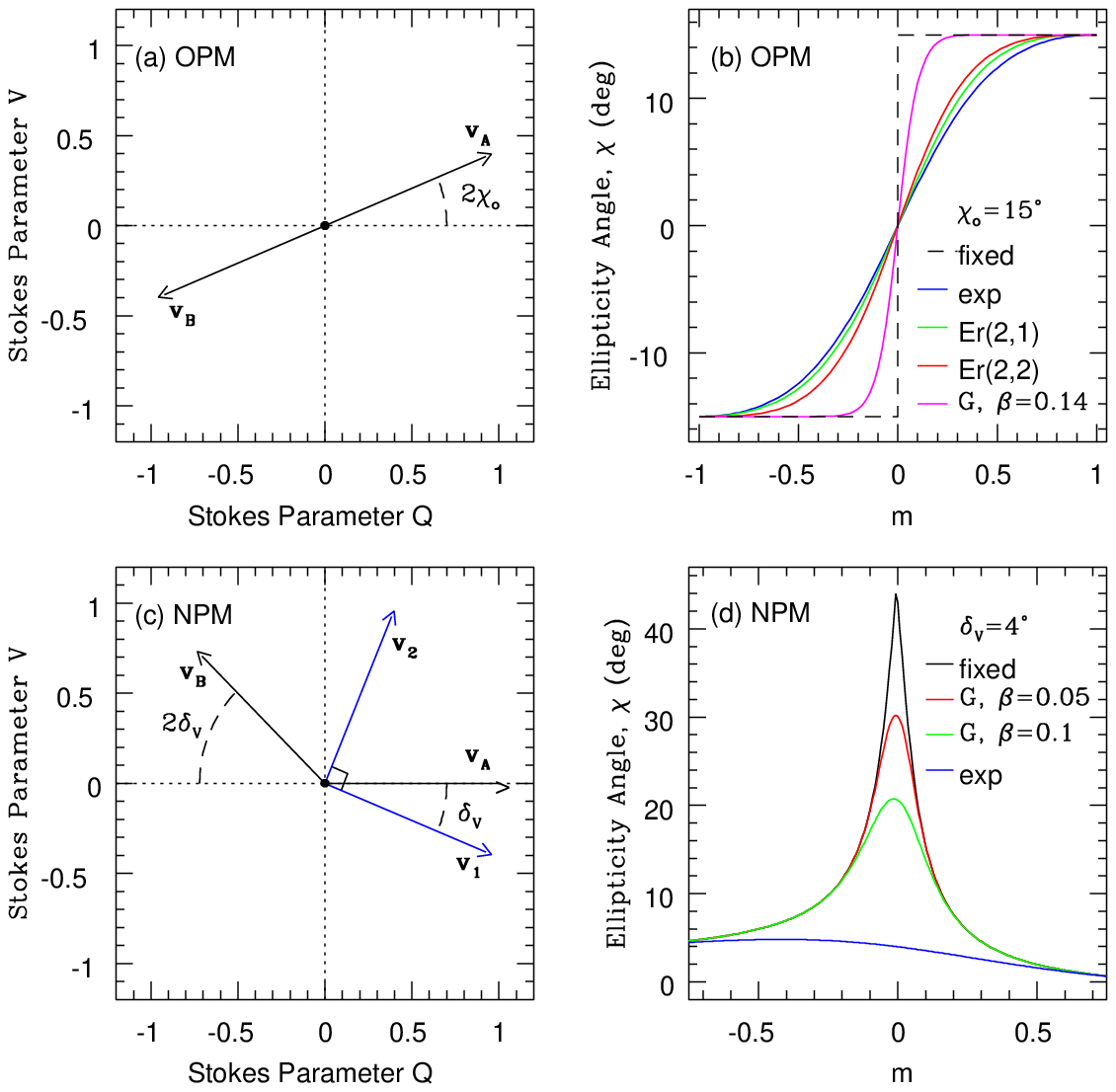}
\caption{Change in the EA at OPM and NPM transitions when the mode intensities are fixed 
or RVs. Panel (a) shows the polarization vectors of the OPMs, $\mathbf{v_A}$ 
and $\mathbf{v_B}$. Panel (b) shows the measured EA at an OPM transition when the mode 
intensities are fixed or exponential (exp), Erlang (Er), or Gaussian (G) RVs. Panel (c) 
compares the unit vectors of the NPMs with two of the eigenvectors of the QUV covariance 
matrix, $\mathbf{v_1}$ and $\mathbf{v_2}$. Panel (d) shows the measured EAs when the NPM  
intensities are fixed or exponential or Gaussian RVs.}
\label{fig:random}
\end{figure}

\subsection{EA from Fluctuations in Incoherent OPM Intensities}
\label{sec:EArandom}

The average EA is calculated from the mean linear polarization, $\langle L\rangle$. Since 
the fluctuations in the mode intensities contribute to $\langle L\rangle$, the measured 
EA also depends upon the statistical character of the mode intensity fluctuations. The EA 
produced by randomly fluctuating OPMs can be determined by selecting a probability 
distribution for the mode intensities, deriving the mean circular and linear polarization 
of the combined radiation, and calculating the EA from Equation~\ref{eqn:chi}. M22 and M24 
derived the mean linear polarization normalized by the mean total intensity, 
$\bar{\rm L}=\langle L\rangle/\langle I\rangle$, for a variety of mode intensity 
probability distributions. Those results can be used to calculate the EA by replacing 
$\langle V\rangle/\langle L\rangle$ in Equation~\ref{eqn:chi} with $\bar{\rm V}/\bar{\rm L}$, 
where $\bar{\rm V}=\langle V\rangle/\langle I\rangle$ is the mean circular polarization 
normalized by the mean total intensity. From Equations~\ref{eqn:I} and~\ref{eqn:V}, and
regardless of the statistical character of the mode intensity fluctuations, the normalized
mean of the circular polarization is always $\bar{\rm V}=m\sin(2\chi_o)$, where $\chi_o$ is 
shown by the OPM vector geometry illustrated in Figure~\ref{fig:random}(a). For OPM 
intensities that are exponential RVs, the normalized mean of the linear polarization is 
Equation~\ref{eqn:px} multiplied by $\cos(2\chi_o)$, and the EA as a function of $m$ is

\begin{equation}
\chi(m) = \frac{1}{2}\arctan{\left[\frac{2m\tan(2\chi_o)}{1 + m^2}\right]}.
\label{eqn:chiE}
\end{equation}

\noindent The same procedure can be used to derive the EA produced by other types of OPM
intensity distributions. Examples for Gaussian and Erlang OPM intensities are listed in 
Appendix~\ref{sec:EAGE}, and are compared with the EAs expected from fixed and exponential 
mode intensities in Figure~\ref{fig:random}(b). When the OPM intensities are fixed, the EA 
changes discontinuously by $\Delta\chi=2\chi_o$ at the mode transition ($m=0$). When the OPM 
intensities are random, the EA changes gradually and continuously from that of one mode to 
the other. The EA transition for Gaussian mode intensities is more acute than the transition 
resulting from exponential mode intensities. Since the modulation index of a Gaussian RV is 
less than an exponential RV's, the figure shows the EA transitions becoming more acute as the 
modulation index decreases. The EAs for the Erlang distributions are shown in the figure to 
illustrate how the EA changes at a transition when the mode intensity distributions are not 
the same. The red curve labeled Erlang(2,2) (Equation~\ref{eqn:chi22}) is the EA resulting 
from both modes having the same Erlang distribution with their orders equal to $2$
($n_a=n_b=2$; see Appendix~\ref{sec:EAGE} for the definition of $n$). The green curve labeled 
Erlang(2,1) (Equation~\ref{eqn:chi21}) is the EA resulting from different Erlang distributions 
with $n_a=2$ and $n_b=1$. The EA transitions are antisymmetric about $m=0$ when the mode 
intensity distributions are identical, and are asymmetric about $m=0$ when the distributions 
are different. More specifically, the behavior of the EA near $m=1$ in the Erlang(2,1) case 
approaches that of the Erlang(2,2) case, and approaches the exponential case, which is 
Erlang(1,1), near $m=-1$.

% --------------------------------------------------------------------------------------------

\subsection{EA from Fluctuations in Incoherent NPM Intensities}
\label{sec:NPMrandom}

The NPM model is modified in this section to illustrate how the mode intensity fluctuations 
affect the EA at a mode transition. Since the EA behavior derived from fixed mode 
intensities can be similar for all four models, as summarized in Section~\ref{sec:compare}, 
the effect of the fluctuations on the EA derived from the other models is likely to be similar.
For NPMs with fluctuating intensities, the total excursion of the EA at a mode transition 
depends upon the statistical character of the mode intensity fluctuations. From 
Equation~\ref{eqn:chi}, for example, $\langle V\rangle$ must exceed $\langle L\rangle$ for 
the magnitude of the EA to exceed $|\chi|=\pi/8$. This can be difficult to produce, because 
the polarization fluctuations contribute to $\langle L\rangle$, but not to $\langle V\rangle$. 
When the mode intensities do not vary or their fluctuations are small, $\langle V\rangle$ can 
be much larger than $\langle L\rangle$, and the EA will be $\chi=\pm\pi/4$, as shown in 
Figure~\ref{fig:nonorthogonal}(d). When the fluctuations are very large, $\langle V\rangle$ 
will likely be much smaller than $\langle L\rangle$, and the EA will vary gradually by 
$\Delta\chi=\delta_V$ across the mode transition. When the fluctuations are moderate, such 
that $\langle V\rangle\approx\langle L\rangle$, the EA will be $|\chi|\approx \pi/8$.

The effect of the NPM fluctuations on the EA can be quantified by repeating the analysis of 
Section~\ref{sec:EA}, where $\delta_L=0$, while allowing the mode intensities to be random 
instead of fixed. For the NPM vector geometry shown in Figure~\ref{fig:random}(c), where 
$\chi_o=0$, and assuming the NPM intensities are exponential RVs, the measured EA varies with 
$m$ according to

\begin{equation}
\chi(m) = \frac{1}{2}\arctan{\left[\frac{2\tan\delta_V(1-m)(1+m\tan^2\delta_V)}
{(1+m^2)(1+\tan^4\delta_V) + 4m\tan^2\delta_V}\right]}.
\label{eqn:EAexp}
\end{equation}

\noindent The transition still occurs at $m_t=-\tan^2\delta_V$, where the EA is $\chi=\delta_V$.

For NPM intensities that are Gaussian RVs, the EA varies with $m$ as

\begin{equation}
\chi(m) = \frac{1}{2}\arctan{\left\{\frac{\tan\delta_V(1-m)}{\sqrt{2(1+\tan^4\delta_V)}}
\frac{\sqrt{\pi}}{\{\beta\exp[-(f(m)/\beta)^2] + f(m){\rm erf}[f(m)/\beta)]\sqrt{\pi}\}}\right\}},
\label{eqn:EAgauss}
\end{equation}

\noindent where $f(m)$ is

\begin{equation}
f(m) = \frac{m + \tan^2\delta_V}{\sqrt{2(1 + \tan^4\delta_V)}}.
\end{equation}

\noindent Panel (d) of Figure~\ref{fig:random} compares the behavior of the EA at a transition 
between NPMs when the mode intensities are fixed and random. The black line in the panel is the 
EA calculated from Equation~\ref{eqn:EAgen} for fixed mode intensities with $\delta_L=0$. The 
red and green lines show the EAs calculated from Equation~\ref{eqn:EAgauss} for Gaussian mode 
intensities with different modulation indices, as indicated in the panel. The blue line in 
the panel is the EA calculated from Equation~\ref{eqn:EAexp} for exponential mode intensities. 
The figure shows the excursions in EA become more suppressed as the intensity modulation index 
increases. 

% --------------------------------------------------------------------------------------------

\subsection{Numerical Simulation of an Incoherent NPM Transition}
\label{sec:simulation}

\begin{figure}
\plotone{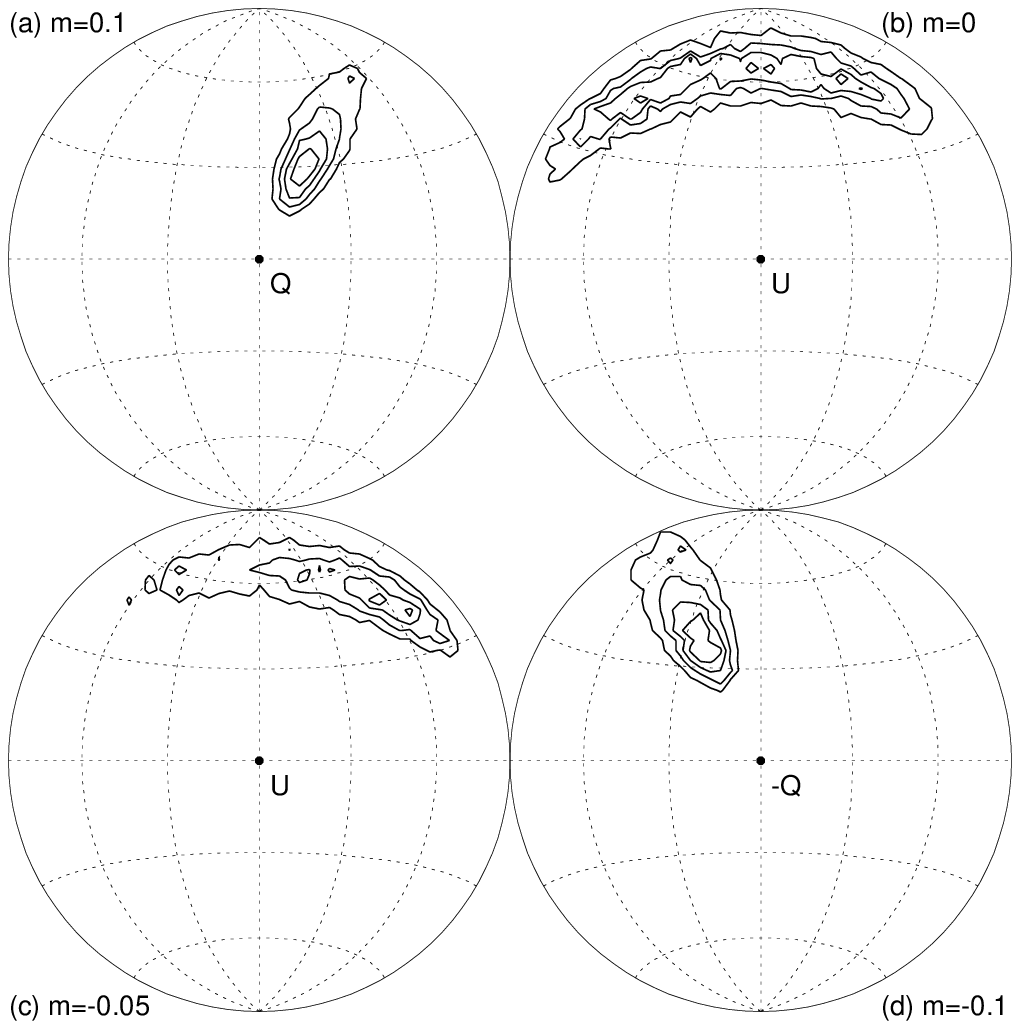}
\caption{Numerical simulations of the trajectory of an NPM transition on the Poincar\'e sphere
when the mode intensities fluctuate. Each panel shows the result of a simulation for different 
values of $m$, as annotated in the figure. The labeled dot at the center of each panel denotes 
the viewpoint of the sphere's projection. The constant angles used to produce the figure are 
$\chi_o=0^\circ$, $\delta_L=3^\circ$ and $\delta_V=6^\circ$. The contour levels are 0.2, 0.4, 
0.6, and 0.8 of the peak value in each projection. See the text for a detailed explanation of 
the simulations.} 
\label{fig:path}
\end{figure}

A numerical simulation was developed to illustrate the effects of mode intensity fluctuations 
upon the trajectory of an NPM transition across the Poincar\'e sphere. The simulation generated 
10,000 samples for each of the Stokes parameters Q, U, and V. The mode intensities were Gaussian 
RVs with identical standard deviations, $\sigma_A=\sigma_B=3$. The mean of the mode B intensity 
was held constant at $\mu_B=30$, while the mean of mode A was varied to produce results at 
different values of $m$. The PA and EA of mode A were held constant at $\psi_A=\chi_A=0^\circ$. 
The PA and EA of mode B were held constant at $\psi_B=\pi/2-\delta_L$, with $\delta_L=3^\circ$, 
and $\chi_B=\delta_V=6^\circ$. From Equation~\ref{eqn:mx}, the mode transition for these values 
of $\delta_L$ and $\delta_V$ occurs at about $m_t=-0.0138$. Instrumental noise was included in 
the simulation with a value of $\sigma_N=1$. The results of the simulation are shown by the 
Lambert equal-area projections in the four panels of Figure~\ref{fig:path} for different values 
of $m$. The most striking feature of all four examples is the data points are not concentrated 
at one location on the sphere, but are instead dispersed along a mean trajectory between the 
orientations of the mode polarization vectors. The mean trajectory is similar to those 
illustrated in Figure~\ref{fig:arcs}. The dispersion of the data points along and about the 
trajectory is caused primarily by the mode intensity fluctuations and, to a much lesser extent, 
by the instrumental noise. The overall trajectory becomes more apparent near the mode 
transition, $m_t$. Panel (a) in the top left corner of the figure shows the result of the 
simulation when $m=0.1$ ($\mu_A=36.67$). The view of the Poincar\'e sphere is along its Q-axis. 
The orientation of the mode A polarization vector is located at the center of the projection. 
As with the other three examples in the figure, the vast majority of data points generated by 
the simulation reside within one hemisphere of the sphere. While mode A is the primary 
(dominant) mode for this value of $m$, the contours are spread about PAs and EAs that are 
significantly different from those of mode A. The contours in the projection trace the 
beginnings of an arc that ventures closer to the sphere's right circular pole than to its 
equator. Panel (b) in the top right corner of the figure shows the result of the simulation 
when $m=0$ ($\mu_A=30$). The view of the sphere in this example is along its U-axis. The 
overall trajectory of the transition is more apparent because the value of $m$ in the 
projection is very close to $m_t$. Mode A remains the primary mode because $m>m_t$. The 
contours also indicate mode A is the primary, because more data points reside on the left 
side of the projection than the right. The average EA is near its maximum in this example. 
Panel (c) in the bottom left corner of the figure shows the result of the simulation when 
$m=-0.05$ ($\mu_A=27.14$). The view of the sphere is again along its U-axis. Mode B is now 
the primary mode because $m<m_t$. The contours follow an arc descending toward the 
orientation of the mode B polarization vector. Panel (d) in the bottom right corner of the 
figure shows the result of the simulation when $m=-0.1$ ($\mu_A=24.54$). The view of the 
sphere is now along its -Q-axis. Mode B is the primary mode for this value of $m$, but the 
contours are spread about PAs and EAs that are significantly different from those of mode B. 
The contours begin to coincide with the orientation of the mode B vector as $m$ approaches 
$m=-1$.

Another numerical simulation was developed to investigate how fluctuations in $\delta_L$ 
and $\delta_V$ might affect the transition trajectory. This modification effectively
enabled multiple trajectories between the orientations of the mode vectors, thereby 
dispersing the polarized signal across the surface of the Poincar\'e sphere and diluting 
the pattern of the arc. The results of this simulation are not shown.

% --------------------------------------------------------------------------------------------

\subsection{Eigenvalues of the Stokes QUV Covariance Matrix}
\label{sec:covariance}

The origin of the polarization fluctuations can be investigated by examining the size and 
shape of the data point cluster formed from multiple samples of the Stokes parameters Q, 
U, and V recorded at a given pulse longitude (McKinnon 2004, hereafter M04). The size and 
shape of the cluster can be quantified by computing the eigenvalues of the QUV covariance 
matrix (M04; ES04). The measured eigenvalues (e.g., Figure 3 of ES04 and Figure 3 of Edwards 
(2004)) can then be compared with the eigenvalues derived from the polarization models.

The eigenvalues for the general case of the NPM model, where departures from orthogonality 
occur in both linear and circular polarization, can be derived by following the analysis 
outlined in Section 2.4 of M04. For consistency, the PAs and EAs of the mode polarization 
vectors used in the derivation are the same as those used in Section~\ref{sec:general}. The 
variances of the mode intensities are assumed to equal one another, 
$\sigma_A^2=\sigma_B^2=\sigma^2$, to simplify the analysis and to focus the result on the 
effects of the vector geometry. The resulting eigenvalues are 

\begin{equation}
\tau_1 = \sigma_N^2 + \sigma^2[1 + \cos(2\delta_V)\cos(2\delta_L)],
\label{eqn:tau1}
\end{equation}

\begin{equation}
\tau_2 = \sigma_N^2 + \sigma^2[1 - \cos(2\delta_V)\cos(2\delta_L)],
\label{eqn:tau2}
\end{equation}

\begin{equation}
\tau_3 = \sigma_N^2,
\label{eqn:tau3}
\end{equation}

\noindent where $\sigma_N$ is the magnitude of the instrumental noise. The eigenvalues are 
generally not equal to one another, indicating the QUV data point cluster does not possess 
an axis of rotational symmetry (M04). The first eigenvalue is larger than the other two. The 
third eigenvalue is the square of the instrumental noise. The second and third eigenvalues 
are equal when the polarization modes are orthogonal ($\delta_L=\delta_V=0$) and are 
approximately equal when the departures from orthogonality are small 
($\delta_L,\delta_V\ll 1$). 

The eigenvectors associated with the eigenvalues are not the same as the mode polarization 
vectors, because the eigenvectors are orthogonal, but the mode vectors are not. Panel (c)
of Figure~\ref{fig:random} compares the unit vectors of the NPMs, $\mathbf{v_A}$ and 
$\mathbf{v_B}$, with two of the eigenvectors of the QUV covariance matrix, $\mathbf{v_1}$
and $\mathbf{v_2}$. The third eigenvector, $\mathbf{v_3}$, is perpendicular to the plane of 
the figure. The panel shows the magnitude of the EA for the first eigenvector is half the
difference between the EAs of the mode A and B vectors. A quantitative comparison between 
the eigenvectors and mode vectors is made in Appendix~\ref{sec:eigenvec}.

% --------------------------------------------------------------------------------------------

\section{Discussion}
\label{sec:discuss}

% ------------------------------------------------------------------------------------

\subsection{EA Excursion by Vector Rotation}
\label{sec:rotate}

D20 produced an EA excursion by rotating a polarization vector across the Poincar\'e sphere. 
He posited a ``polarization patch" located in the equatorial plane of the sphere near, but 
not on, its U-axis (i.e., the DWI mixing angle was $\psi_{mx}\sim\pi/4$). The single patch 
is consistent with his assumption of coherent modes and a polarization vector with fixed 
amplitude and modest signal-to-noise ratio. D20 then rotated the patch about the Q-axis by 
approximately $\pi$ radians. The rotation is equivalent to changing the mode phase lag, 
$\delta_{ox}$, in DWI's COH model. The maximum value of the EA resulting from the vector 
rotation approaches $\chi\sim\pi/4$, and the total change in PA is slightly less than 
$\pi/2$. He then discusses the apparent similarities between the vector rotation and a mode 
transition, but he makes clear that the two mechanisms are not the same. D20 does not cite 
the physical process responsible for the vector rotation, but it is consistent with 
Generalized Faraday Rotation (GFR), where the vector rotates on a small circle of the 
Poincar\'e sphere about a sphere diagonal defined by the mode polarization vectors (Kennett 
\& Melrose 1998). The vector rotation is also consistent with the Faraday pulsation model 
of Cocke \& Pacholczyk (1976). The vector rotation produced by GFR is not limited to $\pi$, 
and can range from 0 to in excess of $2\pi$. The angular radius of the small circle can 
range from 0 to $\pi/2$ ($0\le\psi_{mx}\le\pi/4$), causing the total change in PA to be as 
small as $\Delta\psi=0$ or as large as $\Delta\psi=\pi/2$ and the maximum value of the EA 
to range from 0 to $\pi/4$, depending upon the angular offset between the polarization 
vector and the mode diagonal.

Similarly, OKJ used their PCOH model to show that the EA excursion in PSR J1157-6224 
is consistent with the rotation of the polarization vector at constant $C$ and $R$. The 
coherence factor and mode strength ratio were derived from the near-constant value of the 
polarization fraction at the EA excursion and the maximum value of the EA. Their analysis
showed the mode phase offset, $\eta$, varied linearly with pulse longitude. Both D20 and 
DWI predicted the phase offset might vary in this way. OKJ also found that the EA of PSR 
J0820-1350 changed with wavelength while its polarization fraction remained constant. 
Again assuming that $R$ and $C$ remained constant, they showed the mode phase offset 
varied with wavelength as either $\lambda^2$ or $\lambda^3$. Lower et al. (2024) 
demonstrated that changes in the polarization angles of the radio magnetar XTE J1810-197 
are consistent with a frequency-dependent rotation of its polarization vector. The EPC 
model can also produce a vector rotation by holding $m$ and $\varepsilon$ constant 
while allowing the EA of the EPC, $\chi_e$, to vary.

% ------------------------------------------------------------------------------------

\subsection{Potential Model Applications}

% ------------------------------------------------------------------------------------

\subsubsection{Orientation Angles of the Polarization Vector in PSR B0809+74}

\begin{figure}
\plotone{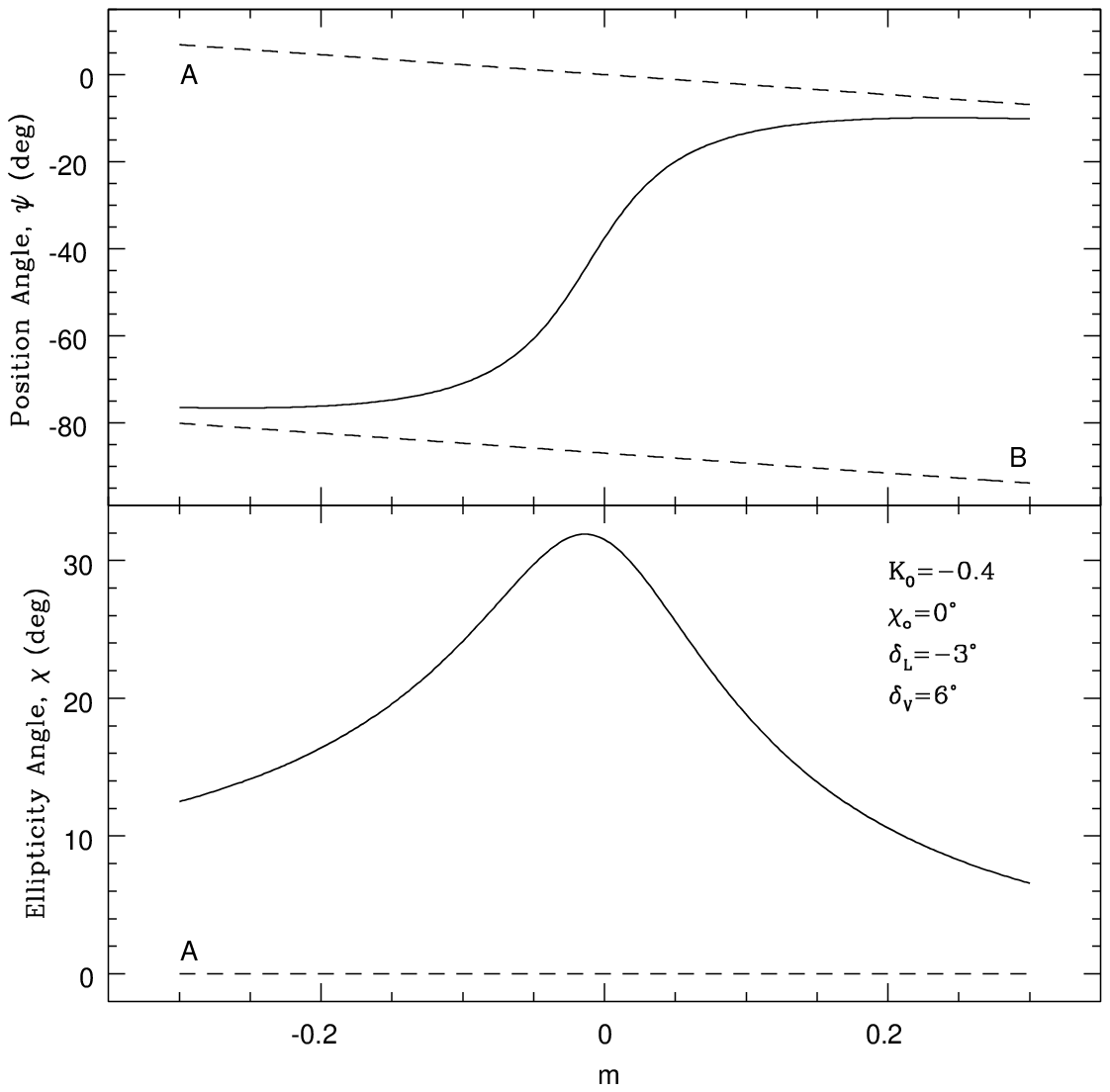}
\caption{Model of the PA and EA in PSR B0809+74 assuming the mode polarization vectors 
are not orthogonal. The solid line in the top panel shows the PA as calculated from 
Equation~\ref{eqn:PAK}. The dashed lines denote the PA traces of the individual polarization 
modes, A and B. The solid line in the bottom panel denotes the EA as calculated from 
Equation~\ref{eqn:EAgen}. The dashed line represents the EA of mode A.}
\label{fig:0809}
\end{figure}

The NPM model is qualitatively consistent with the polarization properties of PSR B0809+74 
at 328 MHz (Figure 2 of Edwards (2004)). The observation shows polarization modes occurring 
across most of its pulse profile. The PA traces of the two modes are approximately parallel 
and vary linearly with negative slopes across the pulse. Ramachandran et al. (2002) analyzed 
the same data set as Edwards (2004) and found the PA separation between the traces was not 
equal to $90^\circ$. The average PA bridges the two traces via a gradual transition near the 
pulse center. The modes are likely responsible for the low polarization observed across the 
pulse. A depression in the polarization is coincident with the PA transition. The average EA 
consists of a broad, asymmetric feature that extends conspicuously over the same range of 
pulse longitude as the PA transition. The feature peaks at $\chi\simeq 32^\circ$ and is 
aligned in pulse longitude with the PA transition. 

Figure~\ref{fig:0809} is an attempt to replicate the observed behavior of the pulsar's PA 
and EA using the equations derived for them in Section~\ref{sec:general}. The solid line in 
the top panel of the figure shows the average PA calculated from Equation~\ref{eqn:PAK}, with 
the PA of mode A varying linearly with $m$, ($\chi_o(m)=K_0m$). The PA traces of the two modes 
are shown by the dashed lines in the panel. The PA trace of mode B is offset from the trace of 
mode A by $\pi/2-|\delta_L|$. The solid line in the bottom panel of the figure shows the average 
EA calculated from Equation~\ref{eqn:EAgen}. The dashed line in the panel denotes the EA of 
mode A, $\chi_o$. The values of $K_0$, $\chi_o$, $\delta_L$, and $\delta_V$ used to calculate 
the PA and EA are annotated in the figure. The qualitative agreement between the calculated and 
observed angles is good. The calculated PA replicates the gradual PA transition that connects 
the PA traces of the individual polarization modes. The calculated EA reproduces the asymmetry, 
longitudinal extent, and peak value of the observed EA.

% ------------------------------------------------------------------------------------

\subsubsection{The Polarization Annulus in PSR B0329+54}

Each model of a mode transition is capable of producing a polarization arc on the Poincar\'e 
sphere, and it is tempting to attribute the partial annulus observed in PSR B0329+54 (Figure 
2 of ES04) to one or more of the models. However, the mode transition interpretation is not 
consistent with the observed annulus, because the models require the orientations of the mode 
polarization vectors to reside at either end of the polarization arc, whereas the symmetry 
axis of the observed annulus appears to be aligned with one of the mode vectors (i.e., the 
mode vectors of the models appear to be perpendicular to the observed mode vectors). ES04 
attribute the annulus to GFR. In its simplest form, GFR requires the modes to be coherent, 
and it would appear as a well-defined arc on the Poincar\'e sphere, with the radial spread 
of the arc determined by the polarization signal-to-noise ratio (e.g., see Figure 3 of 
Kennett \& Melrose 1998). The observed annulus, however, is diffuse, and GFR by itself 
cannot reproduce the compact ellipse observed at the center of the other hemisphere of the 
sphere. Melrose et al. (2006) developed a numerical simulation that reproduced the 
observation. The simulation incorporated superposed NPMs with slightly correlated mode 
intensities and randomly varying orientations. To reproduce both the compact ellipse and the 
annulus, one of the modes had to be completely absent in a significant fraction of pulses. 
Melrose et al. (2006) mention that GFR is complementary to their model, but they stress that 
their model requires the data points within the annulus to be weakly polarized, whereas there 
is no such implication for GFR. The mode transitions produced by the NPM, EPC, and PCOH 
models support their interpretation of a weakly polarized annulus. (The NPM model is 
essentially an analytical version of the NPM transition that is part of their numerical 
simulation; see Figure~\ref{fig:path}(b)). Another interpretation of the partial annulus 
and compact ellipse is a stochastic version of the PCOH model. In this interpretation, the 
compact ellipse would represent the model's incoherent component and the partial annulus 
would represent its coherent component. Since the polarization fraction of the incoherent 
component is independent of $C$ and $\eta$, its position remains fixed on the sphere. 
Fluctuations in the mode intensities, which are equivalent to fluctuations in $R$, cause 
polarization data points to fall within either hemisphere of the Poincar\'e sphere. The 
polarization of the coherent component depends upon the phase offset, $\eta$, and fluctuations 
in $\eta$ will cause its polarization vector to rotate about an axis defined by the 
polarization vector of the incoherent component. The angular extent of the annulus would 
then depend upon the standard deviation of the $\eta$ fluctuations.
 
% --------------------------------------------------------------------------------------------

\section{Summary}
\label{sec:summary}

Four models of pulsar polarization were examined to quantify and understand the behavior of 
a polarization vector's PA and EA at a transition between polarization modes. The results 
of each model show the PA can change gradually, instead of discontinuously, at a mode 
transition, and the EA can trace an excursion that originates near the equator of the 
Poincar\'e sphere and passes near one of its poles. The results of the models can be similar 
to one another, indicating that the interpretation of an observed mode transition within the 
context of a particular model may not be unique. For example, the variations in the 
polarization fractions across a transition that are predicted by the models can resemble one 
another and in some cases are identical. For each model, the trajectory of a transition on 
the Poincar\'e sphere follows a great circle geodesic that connects the orientations of the 
mode polarization vectors. In contrast, the change in PA and EA caused by a vector rotation 
(e.g., via GFR) follows a small circle on the sphere. The COH, PCOH, and EPC models can 
account for changes in the polarization angles due to a vector rotation, in addition to a 
mode transition. Only the NPM model and an enhanced version of the EPC model can produce an 
asymmetric EA excursion about a PA transition, as observed in PSR B0329+54.

The effects of mode intensity fluctuations upon the polarization properties of a mode 
transition were investigated. At a pulse longitude where the mean values of the fluctuating 
mode intensities are approximately equal, individual PA-EA pairs can take on a wide range of 
values constrained by the transition trajectory, forming a polarization arc on the Poincar\'e 
sphere. The polarization fraction and average EA depend upon the statistical character of the 
intensity fluctuations. The polarization fraction increases with the intensity fluctuations. 
An EA excursion can be large when the mode intensities are quasi-stable and is suppressed 
when the intensity fluctuations are large.

% --------------------------------------------------------------------------------------

\acknowledgments{I thank an anonymous referee for constructive comments that improved 
the manuscript. The National Radio Astronomy Observatory is a facility of the National 
Science Foundation operated under cooperative agreement by Associated Universities, Inc.}

% --------------------------------------------------------------------------------------

\appendix

\section{Ellipticity Angles FROM GAUSSIAN AND ERLANG OPM INTENSITIES}
\label{sec:EAGE}

Figure~\ref{fig:random}(b) compares changes in the measured EA at an incoherent OPM 
transition for different types of mode intensity fluctuations. The equations for the 
EAs shown in the figure for Gaussian and Erlang fluctuations are listed in this appendix.

For OPMs with Gaussian intensities, the normalized mean of the linear polarization is 
Equation~\ref{eqn:pg} multiplied by $\cos(2\chi_o)$, and the normalized mean of the 
circular polarization is $\bar{\rm V}=m\sin(2\chi_o)$. The EA from Equation~\ref{eqn:chi} is 

\begin{equation}
\chi(m) = \frac{1}{2}\arctan{\left[\frac{m\tan(2\chi_o)\sqrt{\pi}}
          {\beta\exp(-m^2/2\beta^2)\sqrt{2} 
          + {\rm erf}(m/(\beta\sqrt{2}))m\sqrt{\pi}}\right]}.
\label{eqn:chiG}
\end{equation}

\noindent Equation~\ref{eqn:chiG} is shown by the magenta line in Figure~\ref{fig:random}(b).

The linear polarization resulting from mode intensities that are Erlang RVs was derived in 
M24. For reference, the Erlang distribution is

\begin{equation}
f(x,\mu,n) = \frac{x^{n-1}}{\Gamma(n)\mu^n}\exp{\left(-\frac{x}{\mu}\right)}, \qquad x\ge 0,
\label{eqn:Erlang}
\end{equation}

\noindent where $\mu$ is a positive scaling factor, $n$ is a positive integer representing 
the order of the distribution, and $\Gamma(n)=(n-1)!$ is the gamma function. The general 
equation for $\bar{\rm L}$ for Erlang mode intensities is given by Equation 24 of M24. When 
the orders of the Erlang distributions for OPM intensities are both equal to $2$ ($n_a=n_b=2$), 
the EA is 

\begin{equation}
\chi(m) = \frac{1}{2}\arctan{\left[\frac{8m\tan(2\chi_o)}{3 + 6m^2 - m^4}\right]}.
\label{eqn:chi22}
\end{equation}

\noindent Equation~\ref{eqn:chi22} is shown by the red line in Figure~\ref{fig:random}(b).

For Erlang RVs with $n_a=2$ and $n_b=1$, $\bar{\rm L}$ can be calculated from Equation 26 
of M24. The resulting EA is

\begin{equation}
\chi(m) = \frac{1}{2}\arctan{\left[\frac{m(3-m)^2\tan(2\chi_o)}{4 - 3m(1 - m)^2}\right]}.
\label{eqn:chi21}
\end{equation}

\noindent Equation~\ref{eqn:chi21} is shown by the green line in Figure~\ref{fig:random}(b).

% --------------------------------------------------------------------------------------

\section{EIGENVECTORS OF THE STOKES QUV COVARIANCE MATRIX}
\label{sec:eigenvec}

As noted in Section~\ref{sec:covariance}, the eigenvectors of the Stokes QUV covariance
matrix are generally not the same as the mode polarization vectors. This can be 
demonstrated using the vector geometry shown in Figure~\ref{fig:random}(c), 
where the mode vectors deviate from orthogonality in circular polarization only. The 
unit vectors for the polarization modes shown in the figure are

\begin{equation}
\mathbf{v_A} = \begin{bmatrix} 1 \\ 0 \\  0 \end{bmatrix}, \quad \;
\mathbf{v_B} = \begin{bmatrix} -\cos(2\delta_V) \\ 0 \\  \sin(2\delta_V)\end{bmatrix}.
\end{equation}

\noindent The PA and EA of mode A are $\psi_A=\chi_A=0$. They are $\psi_B=\pi/2$ 
and $\chi_B=\delta_V$ for mode B. The eigenvectors associated with the eigenvalues given by 
Equations~\ref{eqn:tau1} -~\ref{eqn:tau3} when $\delta_L=0$ are

\begin{equation}
\mathbf{v_1} = \begin{bmatrix} \cos\delta_V \\ 0 \\ -\sin\delta_V \end{bmatrix}, \quad \;
\mathbf{v_2} = \begin{bmatrix} \sin\delta_V \\ 0 \\ \cos\delta_V \end{bmatrix}, \quad \;
\mathbf{v_3} = \begin{bmatrix} 0 \\ -1 \\ 0 \end{bmatrix}.
\end{equation}

\noindent The eigenvectors are perpendicular to one another. The first eigenvector, 
$\mathbf{v_1}$, is aligned with the major axis of the QUV data point cluster. Its PA is 
$\psi_1=0$, and its EA is $\chi_1=-\delta_V/2$. The magnitude of $\chi_1$ is half the 
difference between the EAs of modes A and B. When the mode vectors are orthogonal 
($\delta_V=0$), the first eigenvector and the mode A polarization vector are identical.

% -----------------------------------------------------------------------------------


\begin{references}
\reference{} Allen, M. C. \& Melrose, D. B. 1982, Proc. Astron. Soc.  Aust., 4, 365

\reference{} Backer, D. C. \& Rankin, J. M. 1980, \apjs, 42, 143

\reference{} Barnard, J. J. \& Arons, J. 1986, \apj, 302, 138

\reference{} Cheng, A. F. \& Ruderman, M. A. 1979, \apj, 229, 348

\reference{} Cocke, W. J. \& Pacholczyk, A. G. 1976, \apj, 204, L13

\reference{} Cordes, J. M., Rankin, J., \& Backer, D. C. 1978, \apj, 223, 961

\reference{} Dyks, J. 2020, \mnras, 495, L118 (D20)

\reference{} Dyks, J., Weltevrede, P., \& Ilie, C. 2021, \mnras, 501, 2156 (DWI)

\reference{} Edwards, R. T. 2004, \aap, 426, 677

\reference{} Edwards, R. T. \& Stappers, B. W. 2004, \aap, 421, 681 (ES04)

\reference{} Ekers, R. D. \& Moffet, A. T. 1969, \apj, 158, L1

\reference{} Kells, L. M., Kern, W. F., \& Bland, J. R., 1940, Plane and Spherical 
             Trigonometry (New York: McGraw-Hill)

\reference{} Kennett, M. \& Melrose, D. 1998, Proc. Astron. Soc. Aust., 15, 211

\reference{} Lower, M. E., Johnston, S., Lyutikov, M., et al., 2024, \nat, 8, 606

\reference{} Manchester, R. N., Taylor, J. H., \& Huguenin, G. R. 1975, \apj, 196, 83

\reference{} McKinnon, M. M. 2003, \apj, 590, 1026 (M03)

\reference{} McKinnon, M. M. 2004, \apj, 606, 1154 (M04)

\reference{} McKinnon, M. M. 2022, \apj, 937, 92 (M22)

\reference{} McKinnon, M. M. 2024, \apj, 961, 151 (M24)

\reference{} McKinnon, M. M. \& Stinebring, D. R. 1998, \apj, 502, 883 (MS)

\reference{} Melrose, D. B. 1979, Aust. J. Phys., 32, 61

\reference{} Melrose, D. B., Miller, A., Karastergiou, A., \& Luo, Q. 2006, \mnras, 365, 638

\reference{} Oswald, L. S., Johnston, S., Karastergiou, A., et al. 2023a, \mnras, 520, 4961

\reference{} Oswald, L. S., Karastergiou, A., \& Johnston, S. 2023b, \mnras, 525, 840 (OKJ)

\reference{} Petrova, S. A., 2001, \aap, 378, 883

\reference{} Radhakrishnan, V. \& Cooke, D. J. 1969, Astrophys. Lett., 3, 225

\reference{} Ramachandran, R., Rankin, J. M., Stappers, B.W, et al. 2002, \aap, 381, 993

\reference{} Stinebring, D. R., Cordes, J. M., Rankin, J. M., et al. 1984, \apjs, 55, 247 (S84)

\end{references}
\end{document}